\documentclass[aps,prd,fleqn,superscriptaddress]{revtex4}
\pdfoutput=1
\usepackage{graphicx,color,natbib}
\usepackage{amsmath,amssymb,amsfonts}

\usepackage{epsfig,epsf}
\usepackage{dcolumn}
\usepackage{float}
\usepackage{bm}

\newcommand{\bse}{\begin{subequations}}
\newcommand{\ese}{\end{subequations}}
\newcommand{\be}{\begin{equation}}
\newcommand{\ee}{\end{equation}}
\newcommand{\bea}{\begin{eqnarray}}
\newcommand{\eea}{\end{eqnarray}}
\newcommand{\ba}{\begin{array}}
\newcommand{\ea}{\end{array}}

\input amssym.def
\input amssym.tex

\usepackage[colorlinks=true, linkcolor=blue, bookmarks=true]{hyperref}
\begin{document}

\title{Note on stability and holographic subregion complexity}

\author{Mohammad Ali-Akbari\footnote{m\_aliakbari@sbu.ac.ir}}
\affiliation{Department of Physics, Shahid Beheshti University, 1983969411, Tehran, Iran
}
\author{Mahsa Lezgi\footnote{s\_lezgi@sbu.ac.ir}}
\affiliation{Department of Physics, Shahid Beheshti University, 1983969411, Tehran, Iran
}

\begin{abstract}
We study holographic subregion complexity in a spatially anisotropic field theory, which expresses a confinement-deconfinement phase transition. Its holographic dual is a five-dimensional anisotropic holographic model characterized by a Van der Waals-like phase transition between small and large black holes. We propose a new interpretation from the informational perspective to determine the stable and unstable thermodynamically solutions. According to this proposal, the states which need (more) less information to be specified characterize the (un) stable solutions. We similarly offer an interpretation to determine the stable and unstable solutions based on the resource of a computational machine, such that the solutions are (un) stable if computational resource (decreases) increases with the increase of temperature. We observe that the effect of anisotropy on holographic subregion complexity is decreasing. This decreasing effect can be interpreted by considering a whole closed system consisting of the state and its environment in which the complexity of the mixed state decreases and complexity of the environment increases.   
\end{abstract}
\maketitle
\tableofcontents
\section{Introduction}
Gauge-gravity duality or more generally the holographic idea, is an alternative and useful tool for investigation of a strong coupling system, where standard methods are powerless \cite{CasalderreySolana:2011us}. Quark-gluon plasma (QGP) produced in heavy ion collisions presents a strongly coupled fluid with a small viscosity that is an example of such a system \cite{one}. It is believed that the QGP is produced after a very short time after the collision, $\tau_{\rm{therm}}\approx \rm{few}~0.1~\rm{fm/c}$ and there are indications that the anisotropic stage of the QGP takes place for time scales of $0.1~\rm{fm/c}\lesssim \tau\lesssim 0.3-2~\rm{fm/c}$ \cite{two}.

According to gauge-gravity duality, a strongly coupled gauge theory in $d$-dimensional space-time corresponds to a classical gravity in $d+1$-dimensional space-time, therefore, to describe different phenomena in strongly coupled and anisotropic field theories in this case, it suffices to find a proper description on the gravity side. 
 Different effects of anisotropy have been studied in holographic contexts \cite{three,four}. The Lifshitz-like background, which is holographically dual to an anisotropic plasma, is applied to study various aspects of the  anisotropic plasma. For example, in these models, holographic estimates of the total multiplicity for certain values of critical exponent
make what is desirable from experimental data \cite{five}.

One of the interesting areas of this duality is a great connection which is developed to connect the quantities in quantum information theory and certain geometric quantities in the bulk geometry. Complexity is an important concept in quantum information theory. This quantity refers to the time and space resources required to perform a computation efficiently \cite{co1}. In the context of quantum field theory, complexity is defined as the minimum number of unitary operators needed to prepare a target state from a reference state \cite{co2}. In order to describe the complexity holographically, the complexity of pure states in the whole boundary field theory has two proposals namely CV (complexity=volume) conjecture and CA (complexity=action) conjecture. In CA conjecture, the complexity is obtained by the bulk action evaluated on the Wheeler-de Witt patch anchored at some boundary time. In CV conjecture, the complexity is defined as the volume of the extremal/maximal volume of a codimensional-one hypersurface in the bulk ending on a time slice of the boundary \cite{co3,co4}.  
Inspired by the Hubney-Ryu-Takayanagi proposal, the CV proposal extends to be defined on subregions corresponding to the complexity of mixed states, which is known as holographic subregion complexity (HSC), in which the complexity of a subsystem on the boundary equals the volume of codimensional-one hypersurface enclosed by  Hubney-Ryu-Takayanagi surface \cite{alishahiha}. There are a lot of works on CV, CA and HSC for various gravity models in the literature \cite{co5,co6,co7,co8,co9,co10,co11,co12,co13,co14,co15}.    

According to thermodynamics of complexity, complexity is like the entropy of an auxiliary classical system, which describes the evolution of unitary operators \cite{unco}.
For instance, the second law of complexity is the usual second law of thermodynamics applied to the auxiliary classical system. According to this analogy, the difference between the maximum entropy and the actual entropy, which is a resource for doing a work has a parallel definition in the context of complexity known as {\it{uncomplexity}}. It refers to the difference between maximum possible complexity and the actual complexity and in this case work means doing directed computation i.e. a computation with a goal, which is called {\it{computational work}} \cite{unco2}. 

In this paper, in order to study the anisotropy influence on HSC and provide an informational view of the phase transition, we start to review an anisotropic 5-dimensional background specified by an arbitrary dynamical exponent, a nontrivial warp factor and two Maxwell fields that supports the Van der Waals-like phase transition between small and large black holes, in section \ref{1}. Then we calculate HSC in this model for two cases of strip entangling orientation, parallel and perpendicular to the isotropic coordinate, in section \ref{2}. Finally, we present our finding and numerical results in section \ref{3}.

\section{Review on the background}\label{1}
\begin{figure}[H]
\centering
\includegraphics[width=70 mm]{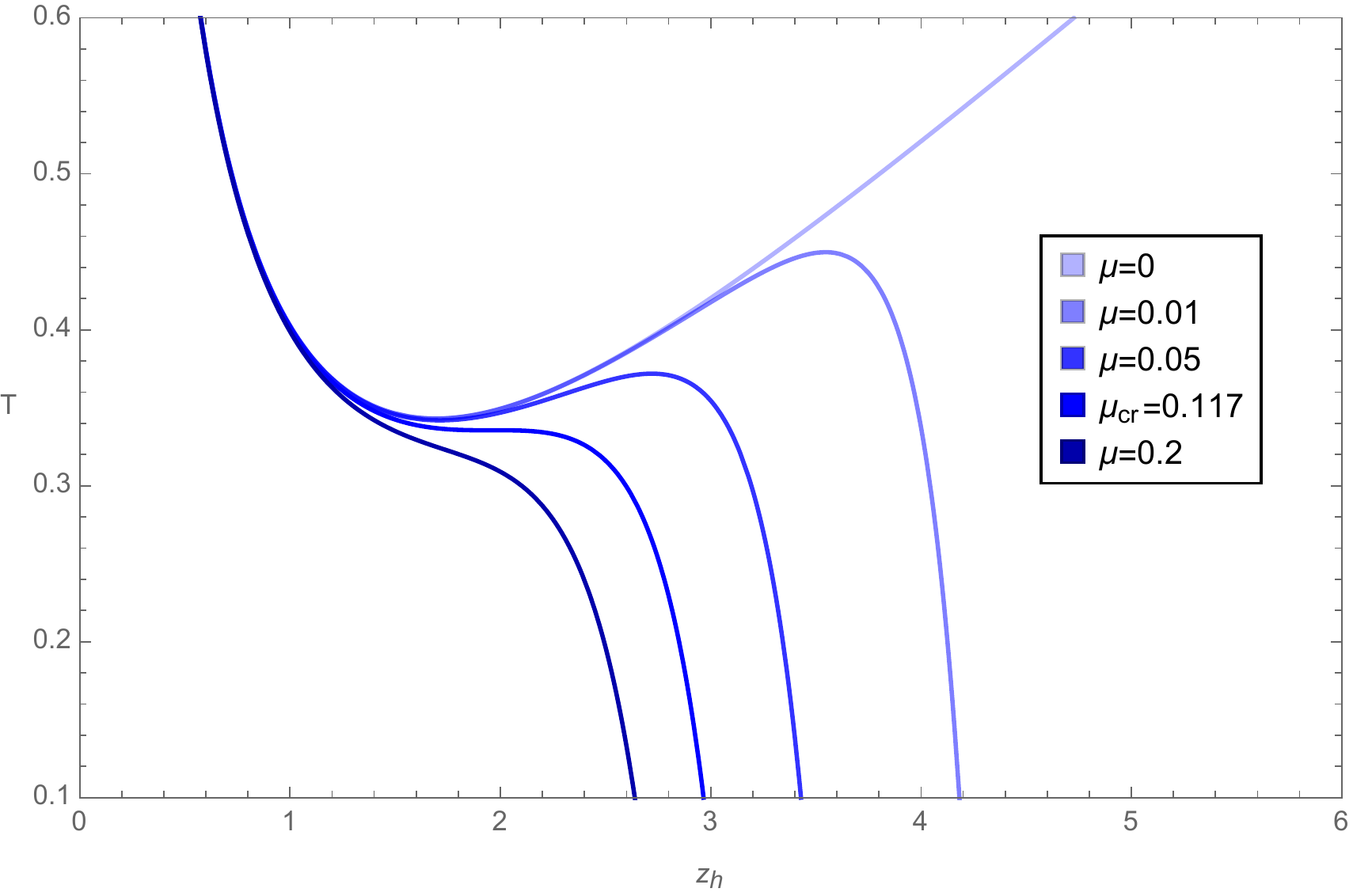} 
\includegraphics[width=70 mm]{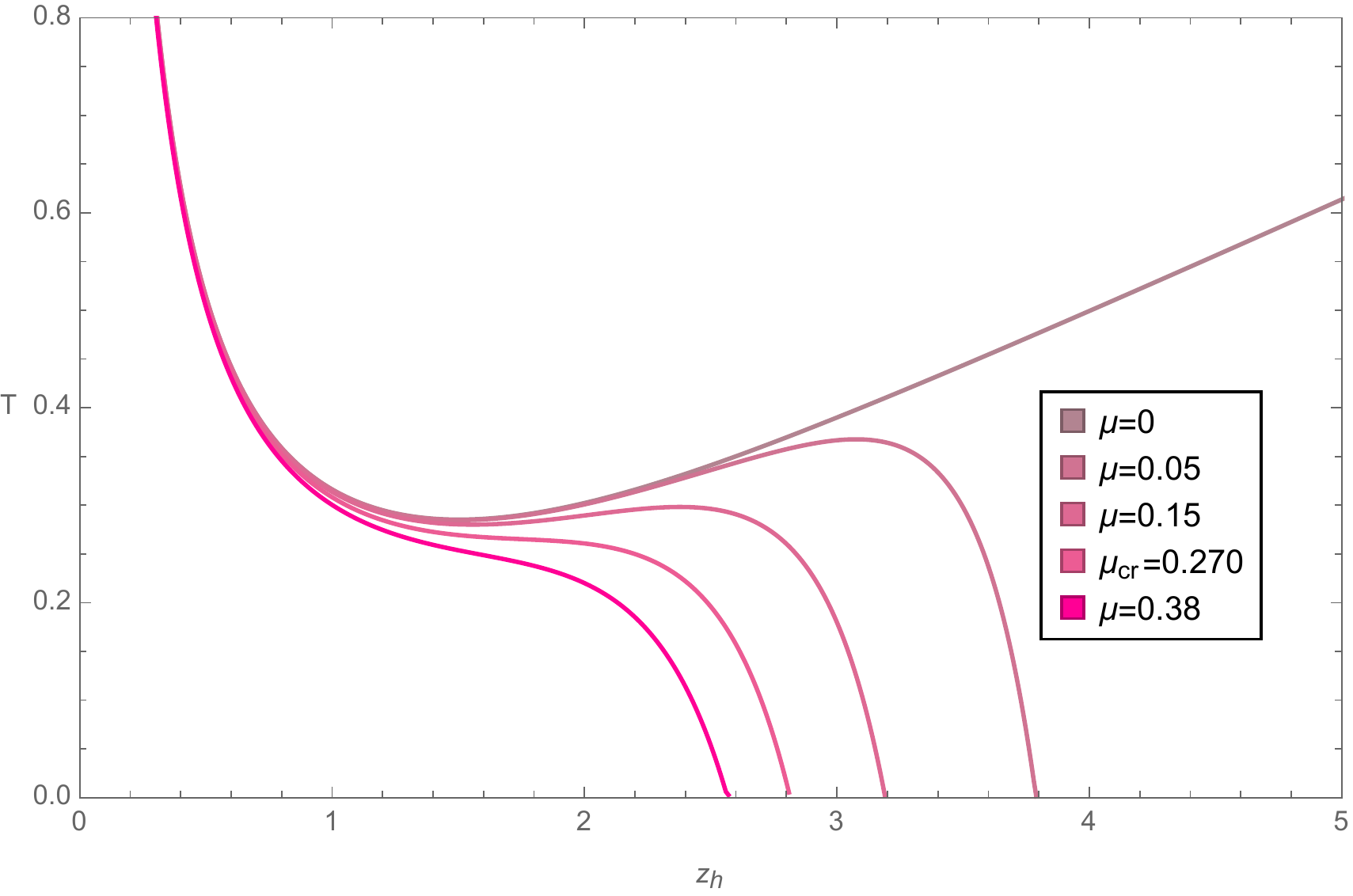} 
\caption{Temperature as a function of the horizon for isotropic case (left) and anisotropic case, $\nu=2$ (right) for different values of $\mu$.}
\label{fig0tem}
\end{figure}

Motivated by the experimental results about the dependence of total multiplicity on energy, anisotropic holographic models related with the Lifshitz-like background, have been studied \cite{ex1,ex2}. The metric we are interested in is supported by a five-dimensional Einstein-Dilaton-two-Maxwell action and introduced in \cite{metric} as follows
\begin{equation}
ds^2=\frac{R^{2}b_{s}(z)}{z^2}\left(-g(z)dt^{2}+dx_{1}^{2}+z^{2-\frac{2}{\nu}}(dx_{2}^{2}+dx_{3}^{2})+\frac{dz^{2}}{g(z)}\right),
\label{metric}
\end{equation}
where $z$ is radial direction and boundary coordinates where the field theory lives on are $t,x_{1},x_{2}$ and $x_{3}$. It is anisotropic along the boundary directions and we call $x_{1}$ and $\nu$, the isotropic coordinate and anisotropic parameter, respectively. $R$ is the characteristic length scale of the geometry that we set as equal to one in our numerical results. $b(z)=e^{cz^{2}/2}$ is the warp factor. Since for $c<0$ the dilaton field becomes real \cite{metric}, in all our calculations we set $c=-1$. The blackening function, $g(z)$, satisfies $g(z_{h})=0$ and $g(0)=1$ in which $z_{h}$ is the position of the horizon, and it must hold the following equation \cite{metric}:
\begin{equation}
g''+g'\left(\frac{3cz}{2}-\frac{1}{z}-\frac{2}{\nu z}\right)-\frac{\mu^{2}c^{2}z^{2+\frac{2}{\nu}}e^{-cz^{2}}}{4\left(1-e^{-\frac{c zh^{2}}{4}}\right)^{2}}=0,
\label{metric2}
\end{equation}
where $\mu$ is the chemical potential in the field theory, due to the existence of the gauge field in the bulk \cite{metric}. The Hawking temperature is given as $T=|g'(z_{h})|/4\pi$. We have plotted the temperature as a function of horizon at different chemical potentials for isotropic (where $\nu=1$) and anisotropic cases in Fig \ref{fig0tem}. In both cases, at $\mu=0$, the temperature has a minimum $T_{\rm{min}}$ at $z_{h}=z_{\rm{min}}$. The branch, corresponding to $0<z_{h}<z_{\rm{min}}$ (large black holes) is thermodynamically stable and the branch in which temperature increases with the horizon, $z_{h}>z_{\rm{min}}$, (small black holes) is unstable \cite{metric}. Below the minimum temperature there is no black hole solution and therefore, there exists a Hawking-Page phase transition where, a thermal AdS solution transits to the large black hole solution at temperature $T\geq T_{HP}$, and $T_{HP}\gtrsim T_{\rm{min}}$. For $0<\mu<\mu_{cr}$, the temperature has a local minimum, $\bar{T}_{\rm{min}}$ and a local maximum, $\bar{T}_{\rm{max}}$ at the values of the horizon, which we call $\bar{z}_{\rm{min}}$ and $\bar{z}_{\rm{max}}$, respectively. The black holes between $\bar{z}_{\rm{min}}$ and $\bar{z}_{\rm{max}}$ are thermodynamically unstable and there is a black hole-black hole phase transition between small black holes corresponding to $z_{h}>\bar{z}_{\rm{max}}$ and large black holes corresponding to $0<z_{h}<\bar{z}_{\rm{min}}$ at $T_{BB}\gtrsim \bar{T}_{\rm{min}}$ \citep{metric}. For $\mu>\mu_{cr}$, the temperature decreases with the growth of the horizon, monotonously and there is no transition anymore. In fact, a second order phase transition happens at the critical point, i.e. $\mu=\mu_{cr}$. Note that as you can see in Fig \ref{fig0tem}, the values of the critical point depend on the values of the anisotropic parameter. To find more properties of this background, we refer the reader to \cite{metric}. To study the potential of quark pairs you can see \cite{metric2} and you can find an investigation of holographic entanglement entropy in this background in \cite{metric3}.       
\section{Holographic subregion complexity}\label{2}
\begin{figure}[H]
\centering
\includegraphics[width=78 mm]{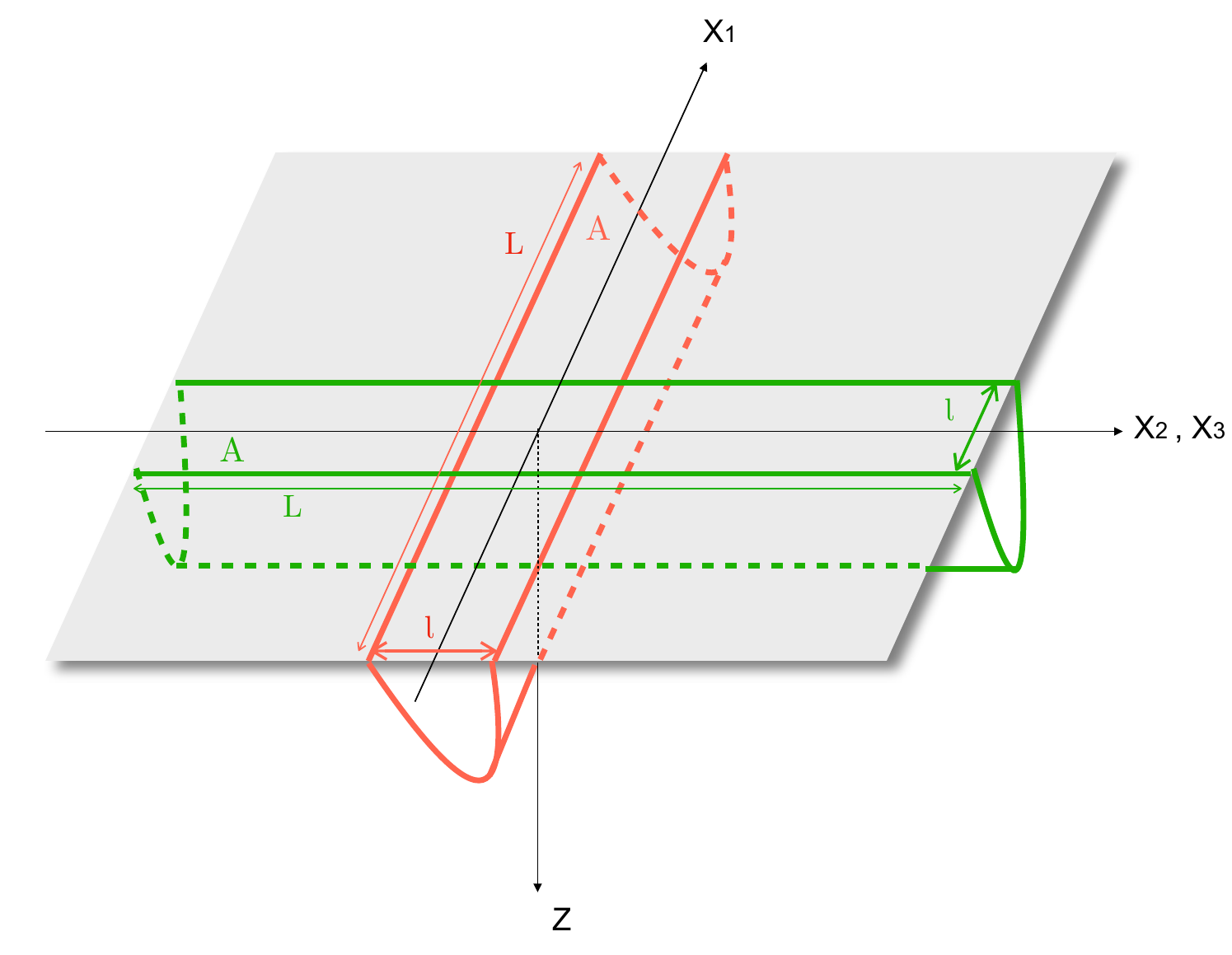} 
\caption{Parallel minimal surface (green) and perpendicular minimal surface (red).}
\label{fig0}
\end{figure}

Inspired by the Hubney-Ryu-Takayanagi proposal, the CV proposal extends to mixed states. HSC cojecture states that the complexity of the mixed state defined on the boundary with a subregion like a rectangular region $A$ in our case, is dual to the volume enclosed by the minimal hyper-surface $\gamma_{A}$, appearing in the holographic entanglement entropy proposal \cite{alishahiha}, i.e. 
\be 
{\cal{C}}_{A}=\frac{V_{\gamma_{A}}}{8\pi R G_{N}},
\label{cv}
\ee
where $R$ and $G_{5}$ are AdS radius and the five-dimensional Newton constant, respectively. ${\cal{C}}_{A}$ is the HSC for the subregion $A$.
The strip entangling surface which is considered here, is defined with one dimension of the length $l$ in $x_{1}$ in parallel to the isotropic coordinate, or in $x_{2}$ or $x_{3}$ (since there is a rotational symmetry in these two directions, there is no difference between them) in perpendicular to the isotropic coordinate. In both cases the surface has a width $L\rightarrow \infty$, see Fig \ref{fig0}. Using \eqref{metric}, we easily find the area of the parallel and perpendicular minimal surfaces as follows
\begin{align}
& S^{||}=\frac{L^{2}}{4G_{5}}\int_{-l/2}^{l/2}\frac{R^{3}b(z)^{\frac{3}{2}}}{z^{3}}z^{2-\frac{2}{\nu}}\sqrt{1+\frac{z'(x_{1})^{2}}{g(z)}}dx_{1},\label{parallel action} \\& S^{\bot}=\frac{L^{2}}{4G_{5}}\int_{-l/2}^{l/2}\frac{R^{3}b(z)^{\frac{3}{2}}}{z^{3}}z^{1-\frac{1}{\nu}}\sqrt{z^{2-\frac{2}{\nu}}+\frac{z'(x_{2})^{2}}{g(z)}}dx_{2},
\label{perpendicular action}
\end{align}
where $z(x_{1})$ (or, equivalently, $x_{1}(z)$) in parallel case and $z(x_{2})$ (or $x_{2}(z)$) in perpendicular case are the profile of the minimal surface. By using the constant of motion, these profiles are obtained:
\begin{align}
&x_{1}(z)=2\int_{z}^{z_{*}}\frac{1}{\sqrt{g(z)}}\left(\frac{b(z)^{3}}{b(z_{*})^{3}}\frac{z_{*}^{2+\frac{4}{\nu}}}{z^{2+\frac{4}{\nu}}}-1\right)^{-\frac{1}{2}}dz, \label{profile1} \\& x_{2}(z)=2\int_{z}^{z_{*}}\frac{1}{\sqrt{g(z)}}\left(\frac{b(z)^3}{b(z_{*})^{3}}\frac{z_{*}^{2+\frac{4}{\nu}}}{z^{\frac{6}{\nu}}}-z^{2-\frac{2}{\nu}}\right)^{-\frac{1}{2}}dz, \label{profile2}
\end{align}
where $z_{*}$ is the turning point of the minimal surface and lies at $x_{1}=0$ in parallel case and $x_{2}=0$ in perpendicular case. In this static background, by slicing the bulk with planes of constant $z$, the volume enclosed by $\gamma_{A}$ is obtained by integrating the inside of the parallel minimal surface as follows 
\begin{equation}
V(z_{*})^{||}=2L^{2}\int_{0}^{z_{*}}\frac{R^{4}b_{s}(z)^{2}}{z^{4}}z^{2-\frac{2}{\nu}}\sqrt{\frac{1}{g(z)}}x_{1}(z)dz.
\label{volume}
\end{equation}
$V(z_{*})^{\bot}$ is obtained by replacing $x_{1}(z)$ with $x_{2}(z)$ in equation \eqref{volume}. Putting anisotropic parameter $\nu=1$ in equation \eqref{volume}, leads to $V_{iso}$. In this case, we have $x_{1}(z)=x_{2}(z)$ and $V^{||}=V^{\bot}=V_{iso}$. 

Volumes similar to the case of entanglement entropy are divergent. We introduce a normalized volume using \eqref{cv} for isotropic and anisotropic cases, respectively,  
\begin{align}
&C_{iso}\equiv\frac{8\pi RG_{5}}{L^{2}}(\mathcal{C}_{iso}-\mathcal{C}_{AdS})=\frac{V_{iso}-V_{AdS}}{L^{2}}, \label{normal1} \\& C^{||or\perp}\equiv\frac{8\pi RG_{5}}{L^{2}}(\mathcal{C}^{||or\perp}-\mathcal{C}_{0}^{||or\perp})=\frac{V^{||or\perp}-V_{0}^{||or\perp}}{L^{2}}. \label{normal2}
\end{align}
In equation \eqref{normal1}, $\mathcal{C}_{iso}$ and $\mathcal{C}_{AdS}$ are the HSC for $A$ in isotropic form of \eqref{metric} and AdS background, respectively. The volumes are defined for the same boundary region such that $V_{iso}$ reduces to $V_{AdS}$ by setting $c$ and $\mu$ equal to zero and $g(z)=1$.
In equation \eqref{normal2}, $\mathcal{C}^{||or\perp}$ and  $\mathcal{C}_{0}^{||or\perp}$ are the HSC for $A$, parallel or perpendicular volumes in the anisotropic case, in the metric \eqref{metric} and anisotropic geometry with the same value of $\nu$ and $g(z)=1$ (zero temperature solution), respectively.  
\section{Numerical results}\label{3}
In this section, we will present our findings from the numerical calculations of normalized HSC, which is introduced in equations \eqref{normal1} and \eqref{normal2} in isotropic and anisotropic cases.
\subsection{Isotropic background}
\begin{figure}[H]
\centering
\includegraphics[width=70 mm]{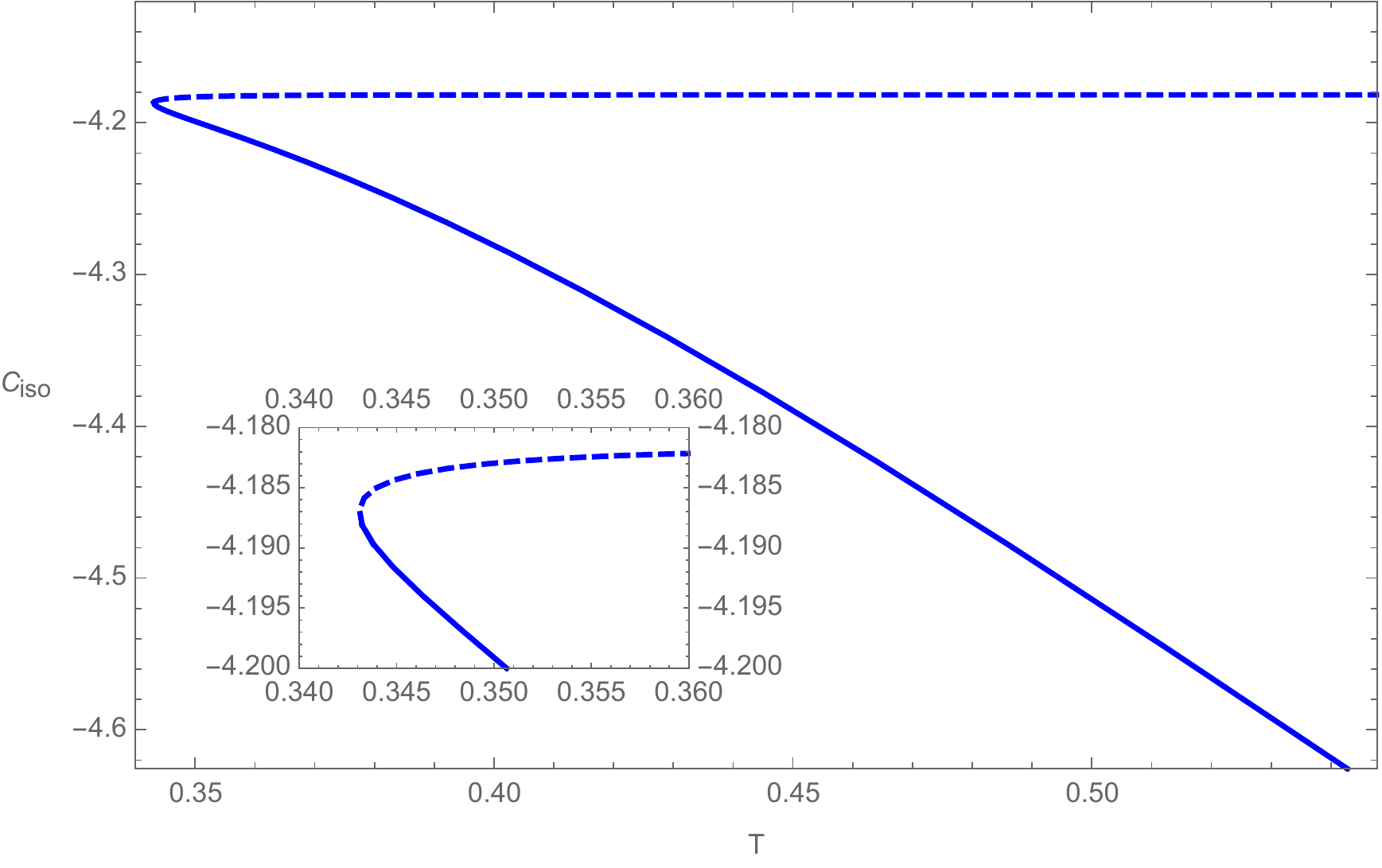}
\includegraphics[width=70 mm]{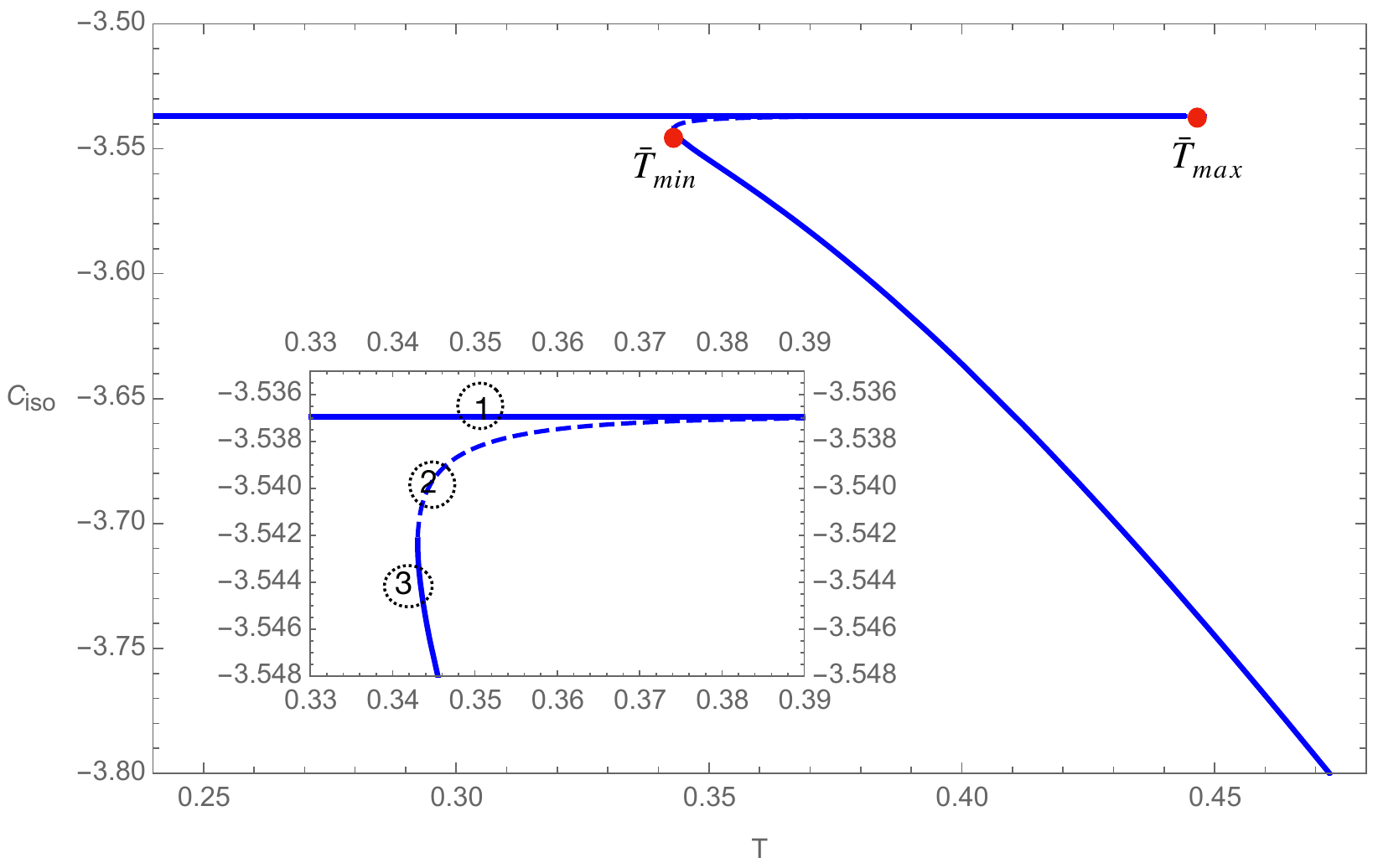}
\caption{left: $C_{iso}$ as a function of temperature for $\mu=0$ and $l=0.5$. Right: The dependence of $C_{iso}$ on temperature at fixed $\mu=0.01$ and $l=0.5$.}
\label{fig1}
\end{figure}

In the isotropic background i.e. when we set the anisotropic parameter equal to one, there is a rotational symmetry in three boundary spatial directions. We would like to investigate whether the thermodynamically stable and unstable solutions can be interpreted from a quantum information point of view. To achieve this purpose, we study the behavior of $C_{iso}$ in terms of temperature. In the left panel of Fig \ref{fig1}, $C_{iso}$ has been plotted as a function of temperature for $\mu=0$. This figure shows that $C_{iso}$ is a double-valued function, the large black holes branch and the small black holes one, as expected. In comparison to the picture of temperature as a function of horizon in the left panel of Fig \ref{fig0tem}, the dashed branch corresponds to the small black holes (or thermodynamically unstable solutions) and the thick part corresponds to the large black holes (thermodynamically stable solutions). Interestingly, between two solutions with the same temperature, the solution with smaller value of $C_{iso}$ belongs to the thermodynamically stable branch. According to \eqref{normal1}, the smaller value of $C_{iso}$ means that the amount of information it takes to specify the considered state is less. Therefore, at fixed temperature, the states described with less information, characterize the stable branch. Note that $C_{iso}$, which is the difference between HSC for the non-conformal background and conformal background is always negative, indicating that non-conformal effects decrease HSC in this model.    
It should be noted that there is a maximum value of HSC in the system we investigate. In the left panel of Fig \ref{fig1}, this maximum value corresponds to the value of $C_{iso}$ at $T_{min}\approx T_{HP}$, where a phase transition takes place from the thermal AdS background to the large black hole solution. On the other hand, considering the  black hole as a computational machine, there is a resource for performing a computation work, which is defined as the difference between the maximum possible complexity and the actual complexity \citep{unco2}. In the stable branch (thick line), with the increase of temperature, $C_{iso}$ deviates from this maximum value. This means that the actual complexity of the stable branch tends to deviate from the maximum possible value of complexity with the increase of temperature. Therefore, in the stable branch resource increases with the increase of temperature and in other words we have a better computational machine.
    
As shown in the right panel of Fig \ref{fig1}, at fixed $\mu=0.01$, between $\bar{T}_{\rm{min}}$ and $\bar{T}_{\rm{max}}$, $C_{iso}$ is a multivalued function of temperature with three branches. In the branch labeled with 1, corresponding to the small black holes branch, $C_{iso}$ is constant with the increase of temperature until $\bar{T}_{\rm{max}}$. The thermodynamically unstable branch is displayed as a dashed line until $\bar{T}_{\rm{min}}$. The second stable branch corresponding to the large black holes is labeled with 3. Near $T_{BB}\gtrsim \bar{T}_{\rm{min}}$, as temperature goes up, phase transition happens from the small black holes solutions to the large black holes solutions. It corresponds to a jump from the first stable branch to the second one, $(1\rightarrow 3)$, corresponding to the smaller value of $C_{iso}$. This means the states identified with less information are more favorable thermodynamically. In short, in cases of $\mu=0$ and $\mu\neq 0$, the state that is favorable thermodynamically, needs less information to be specified from the quantum information perspective. Note that we started from the small black hole branch and argued about what happens with rising temperature.
\begin{figure}[H]
\centering
\includegraphics[width=70 mm]{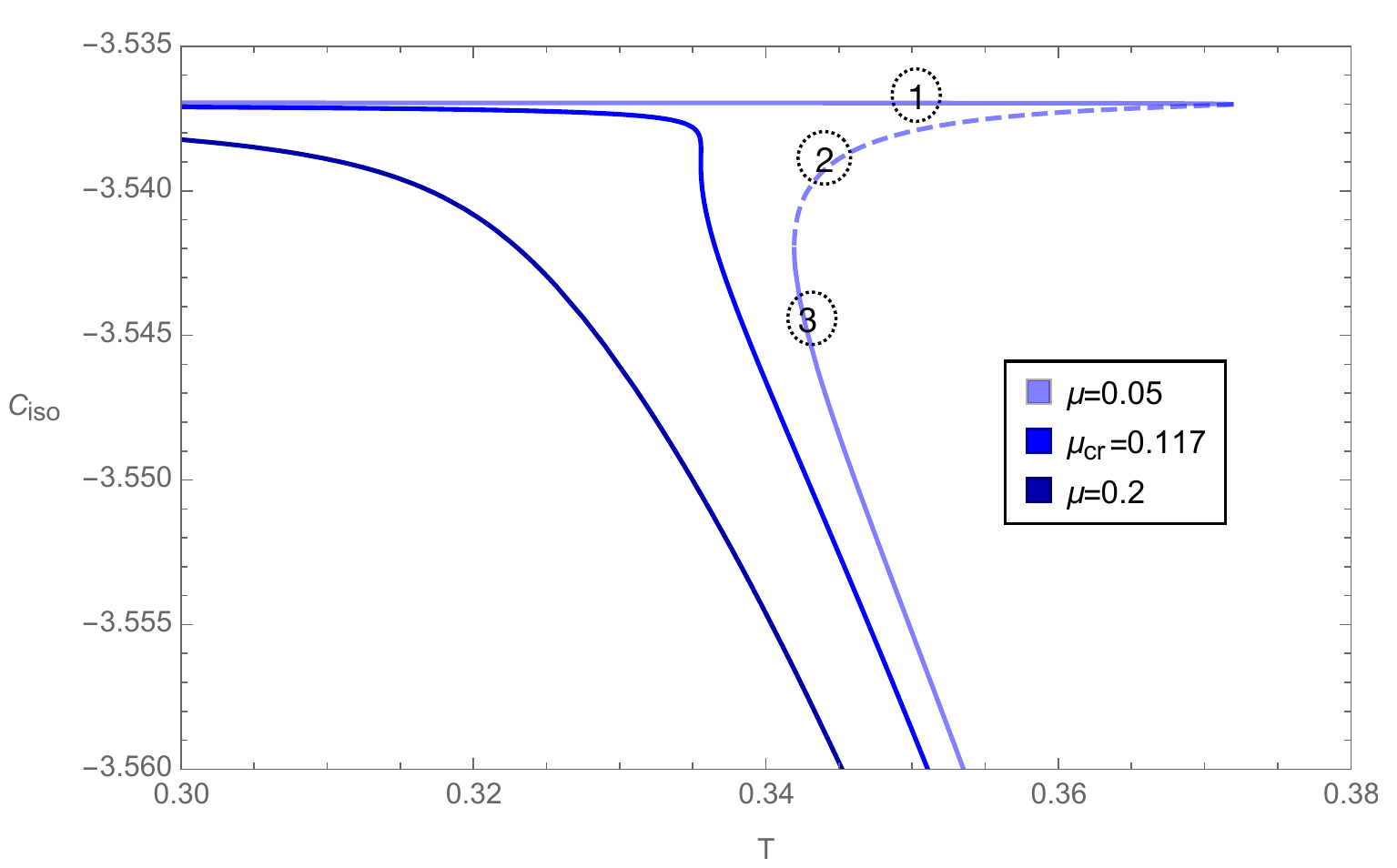}
\caption{$C_{iso}$ in terms of temperature at $l=0.5$ for three values of $\mu$.}
\label{fig2}
\end{figure}  

We want to consider the case that we do not have the assumption mentioned at the end of the previous paragraph i.e we do not start from the small black hole branch. In fact, we are interested in characterizing stable and unstable solutions between the three branches. To reach this goal, we can use the concept of resource. As is shown by the right panel of Fig \ref{fig1}, there are many states with constant maximum value of HSC corresponding to the small black hole branch. It seems that in the presence of $\mu$, as you can see in Fig \ref{fig2}, if $\frac{\mu}{T}$ is small enough, there exist states whose HSC is maximum and independent of the value of $\frac{\mu}{T}$. However, with the increase of $\frac{\mu}{T}$, $C_{iso}$ depends on $\frac{\mu}{T}$. Due to the existence of this maximum, we can apply the interpretation of a computational machine with its resource for doing a computational work. Between these three branches the smaller value of $C_{iso}$ is not enough to determine the stable and unstable solutions. It is because, as shown in Fig \ref{fig2}, the branch labeled 2, which is an unstable solution has less $C_{iso}$ than branch number 1. However, based on definition of resource, in branch number 2 (the unstable branch), the actual complexity is nearing the maximum possible value with the increase of temperature. This means its resource as a computational machine, which is the difference between actual complexity and maximum possible complexity, decreases and thus we have a worse machine for computing with the increase of temperature. Therefore, from the information theory point of view, the solutions are (un) stable if with the increase of temperature, its resource (decreases) increases. As it is summarized in table \ref{list}, zero and increasing resource (branches number 1 and 3) corresponded to stable solutions, while decreasing resource (branch number 2) characterizes the unstable solution.

In Fig \ref{fig2}, we have plotted $C_{iso}$ as a function of temperature for different values of $\mu$. It shows that by nearing $\mu_{cr}$, the multivalued function tends to be the unified one-value function. The dependence of $C_{iso}$ on temperature at this unified branch corresponding to $\mu>\mu_{cr}$ is presented.
\begin{figure}[H]
\centering
\includegraphics[width=70 mm]{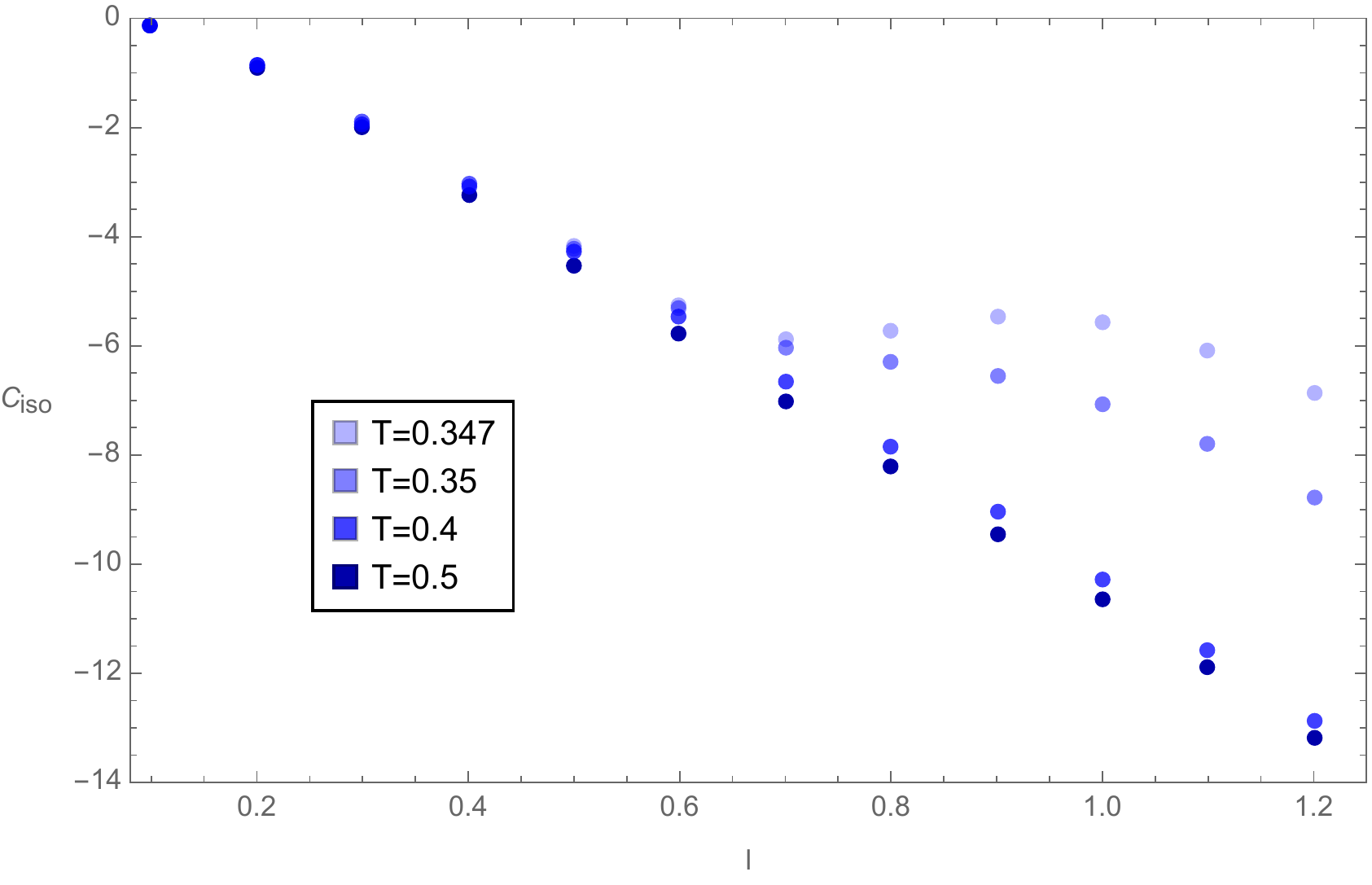}
\includegraphics[width=70 mm]{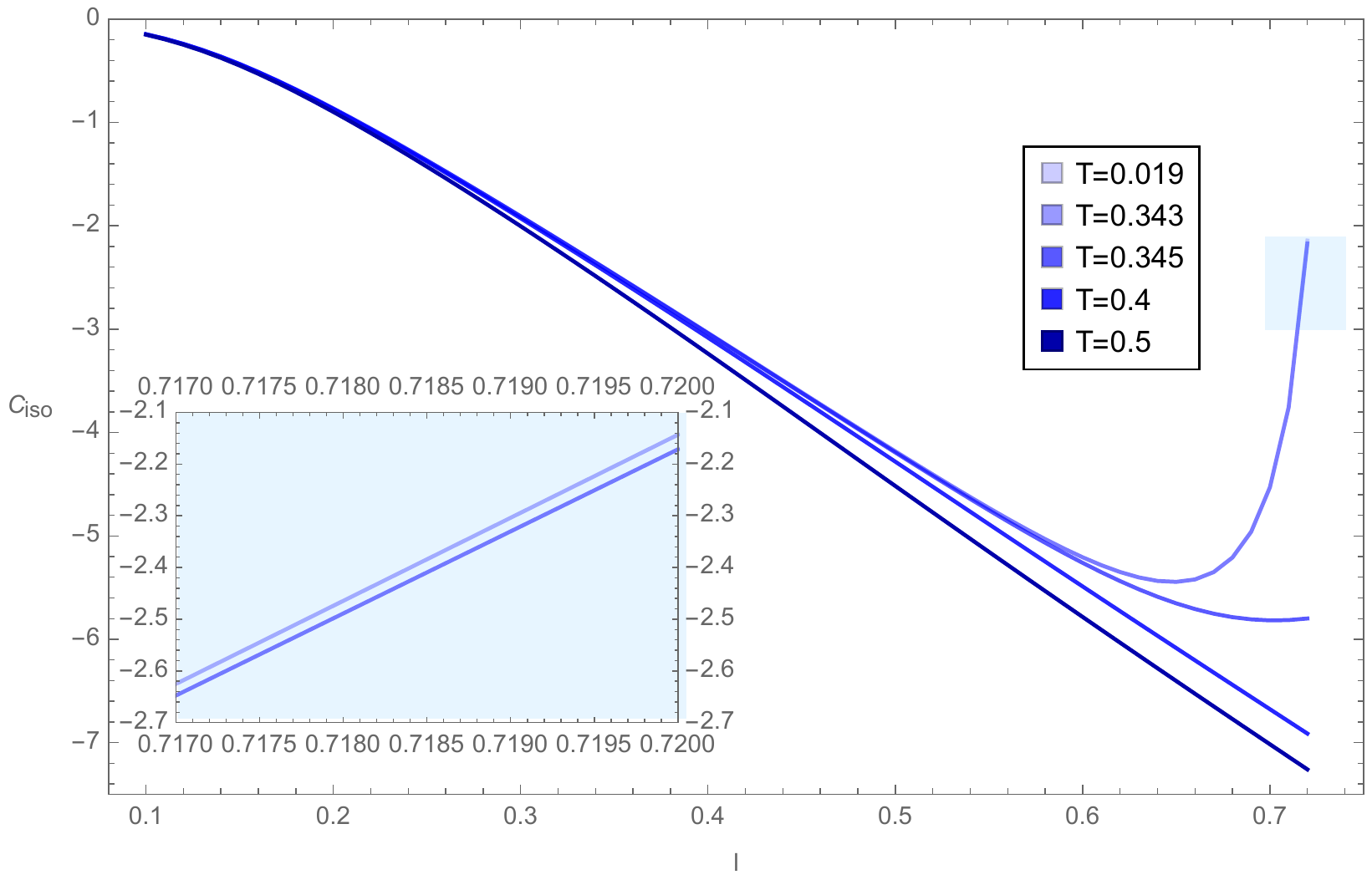}
\caption{Left: $C_{iso}$ in terms of $l$ for $\mu=0$ and four values of temperatures near to $T_{HP}\approx 0.346$. Right: $C_{iso}$ as a function of $l$ for $\mu=0.05$ and different values of the temperature below and above the phase transition temperature, $T_{BB}\approx 0.344$.}
\label{fig3}
\end{figure}  
The left panel of Fig \ref{fig3} shows the behavior of $C_{iso}$ as a function of the subregion length, $l$ for $\mu=0$. This figure has been plotted for the stable branch. For small values of $l$, $C_{iso}$ is almost independent of temperature. However, for larger values of $l$ the difference between values of $C_{iso}$ grows with varying temperatures. In fact, with the increase of $l$, the energy of the probe becomes comparable with $T_{HP}$ and $C_{iso}$ gives more information about the phase transition point. As this Fig shows, at fixed $l$, by comparing non-conformal states, the closer to the phase transition point, the more information needed to specify the state. Moreover, nearing the phase transition temperature $T_{HP}$, this increase changes more with varying temperatures.

In the right panel of Fig \ref{fig3}, we have plotted $C_{iso}$ in terms of $l$ at $\mu=0.05$ for values of the temperature below (the small black hole branch) and above (the large black hole branch) the phase transition temperature. There is the same behavior with $\mu=0$ case for large black hole branches. As you can see, by passing the phase transition point and jumping to the small black hole branch, in a large enough $l$, there is a significant increase in the value of $C_{iso}$ or equivalently, an increase in the needed information to specify the non-conformal state. Another point is that in the small black hole branch for large enough values of $l$, the probe reacts to nearing the phase transition point by increasing the value of $C_{iso}$ and regardless of how far or near the temperature is to the phase transition.

\begin{table}[ht]
\caption{A short summary of resource and stability}
\vspace{1 mm}
\centering
\begin{tabular}{c c c c}
\hline\hline
~~$ \rm{Branch} $ ~~   &~~$ \rm{Resource} $ ~~   &   ~~ $ \rm{Stability} $ ~~  \\[0.5ex]
\hline
$1$ & $\rm{vanishes}$ &  \rm{stable}  \\
$2$ & $\rm{decreasing}$ & \rm{unstable}   \\
$3$ & $\rm{increasing}$ &  \rm{stable}  \\
\hline
\end{tabular}\\[1ex]
\label{list}
\end{table}

\subsection{Anisotropic background}
Here we consider the anisotropic background \eqref{metric} for different values of $\nu$ in two cases, parallel and perpendicular to the isotropic coordinate as shown in Fig \ref{fig0}. The numerical results have been presented as follows.
\subsubsection{Parallel case}
\begin{figure}[H]
\centering
\includegraphics[width=70 mm]{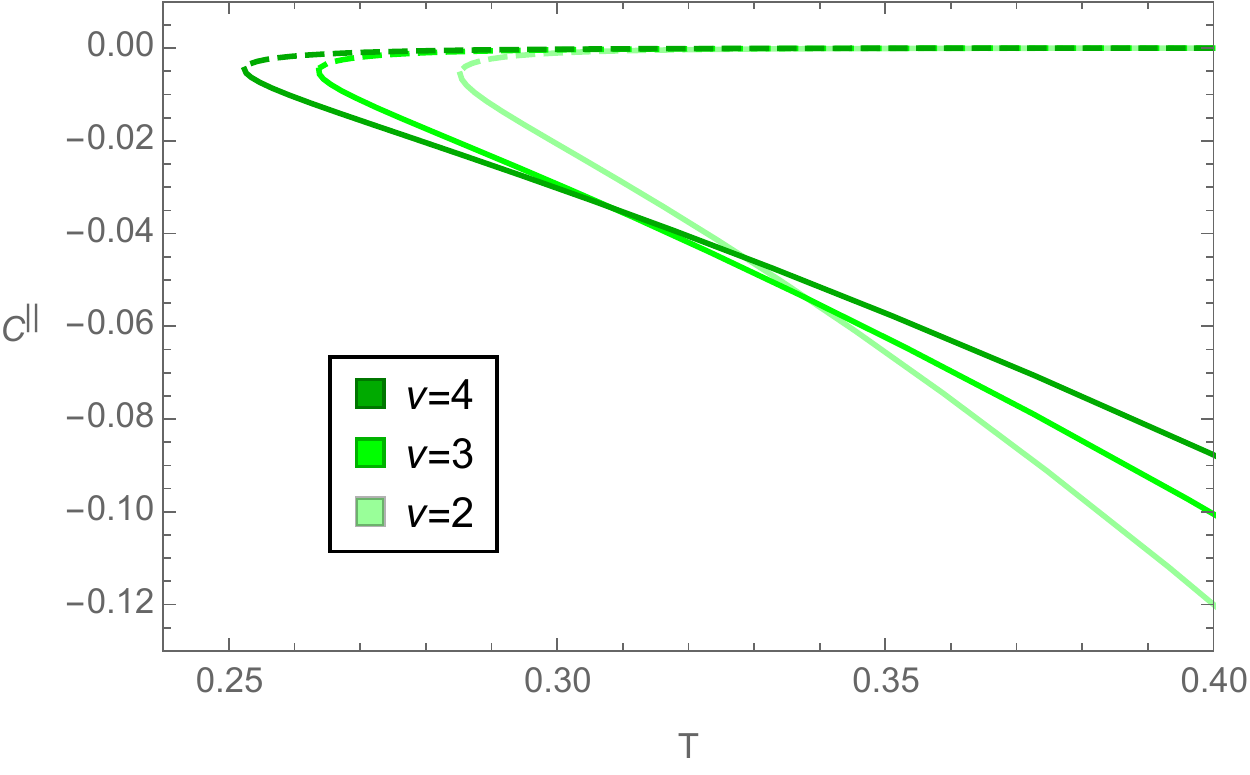}
\includegraphics[width=70 mm]{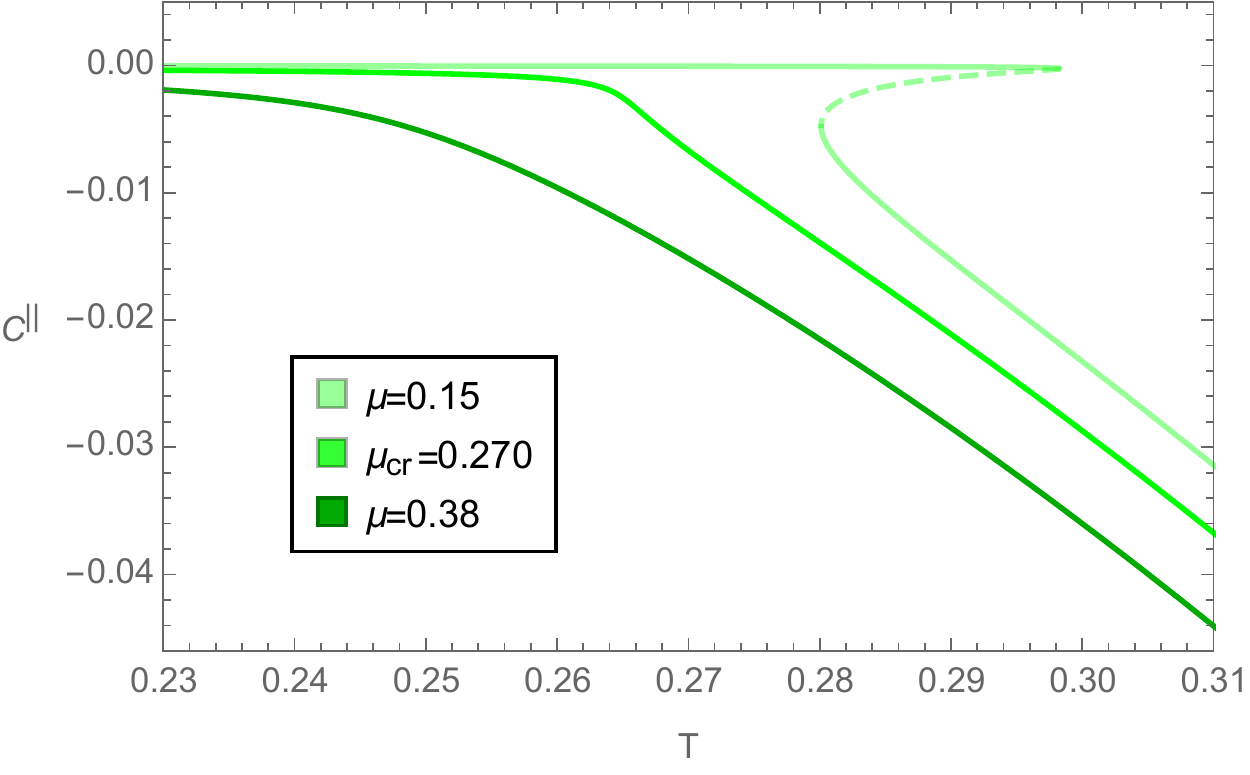}
\caption{Left: $C^{||}$ as a function of $T$ for three values of $\nu$ for $l=0.5$ at $\mu=0$. Right: $C^{||}$ as a function of $T$ for three values of $\mu$ at $l=0.5$ and $\nu=2$.}
\label{fig4}
\end{figure}

Using \eqref{normal2}, we have plotted $C^{||}$ as a function of temperature for different values of $\mu$ in Fig \ref{fig4}. The left panel corresponding to the case of $\mu=0$ shows the thermodynamically stable branch (thick branch) is corresponding to the states with smaller value of $C^{||}$, i.e. ones that need less information to be specified. Hence, like the isotropic case, the states with less required information to be specified characterize the stable branch. Again, there is a maximum value of HSC of the mixed state under study at $T_{min}\approx T_{HP}$ and in the stable branch HSC deviates from this maximum value with the increase of temperature. From the point of view of a computational machine, it means that resource increases and we have a better machine for computing with growing temperature. The left panel shows that with the increase of $\nu$, the minimum of temperature and consequently $T_{HP}$ decreases.
The right panel of Fig \ref{fig4} for the case of $\mu\neq 0$, shows $C^{||}$ as a function of $T$ is a multivalued function between corresponding $\bar{T}_{min}$ and $\bar{T}_{max}$, which is shown in the right panel of Fig \ref{fig0tem}, for $\mu<\mu_{cr}$ in the anisotropic case.  Passing $\mu_{cr}$, this function tends to be a one-value function as expected. In this panel just like the previous section, we can characterize the stable and unstable branches from the information view point.
It is noteworthy that as you can see from Fig \ref{fig4}, $C^{||}$ is negative. According to \eqref{normal2} it  means the effect of temperature on HSC is decreasing. In agreement with \cite{co15}, we can show that our previous argument about this behavior based on an ensemble of microstates corresponding to a given mixed macrostate is valid again. At zero temperature there exists one microstate (unique configuration) but with increasing temperature, the number of microstates corresponding to the mixed macrostate increases. Thus, we specify the state with less information than we do at zero temperature. 

\begin{figure}[H]
\centering 
\includegraphics[width=57 mm]{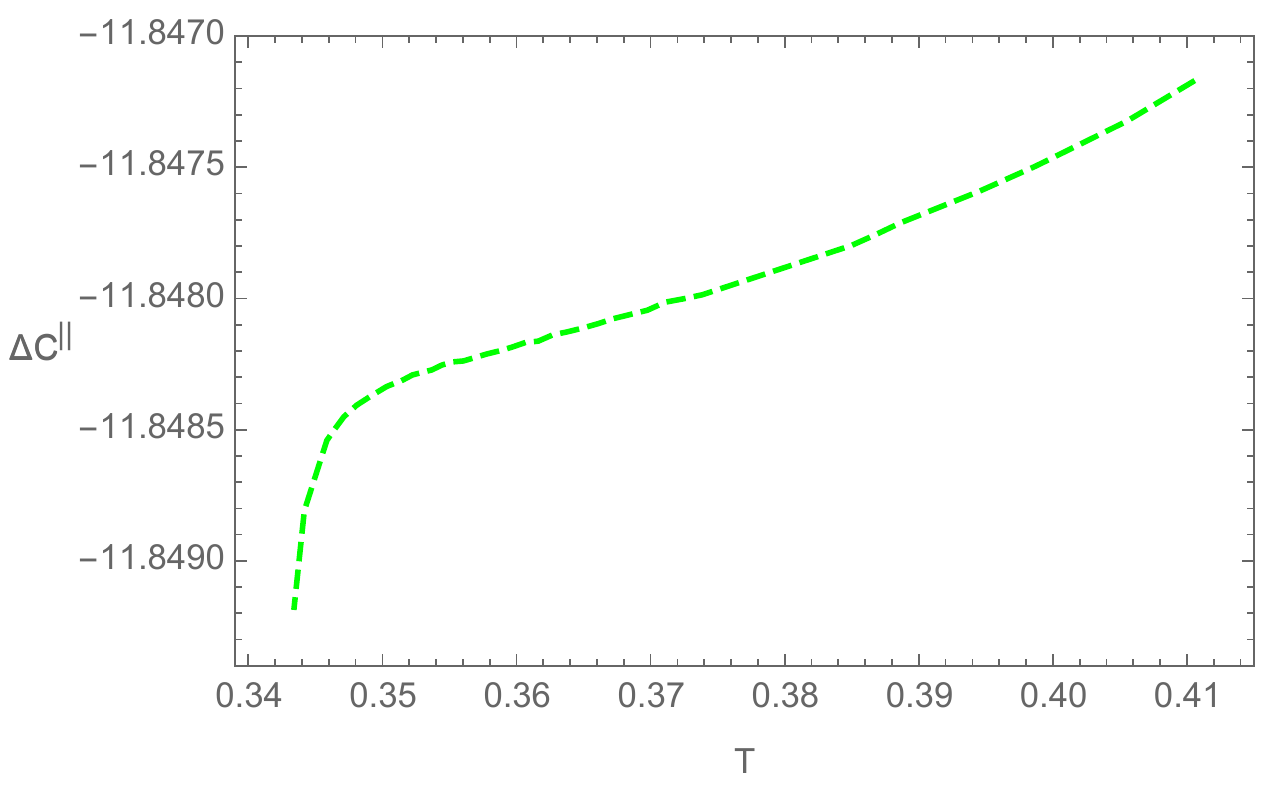}
\includegraphics[width=57 mm]{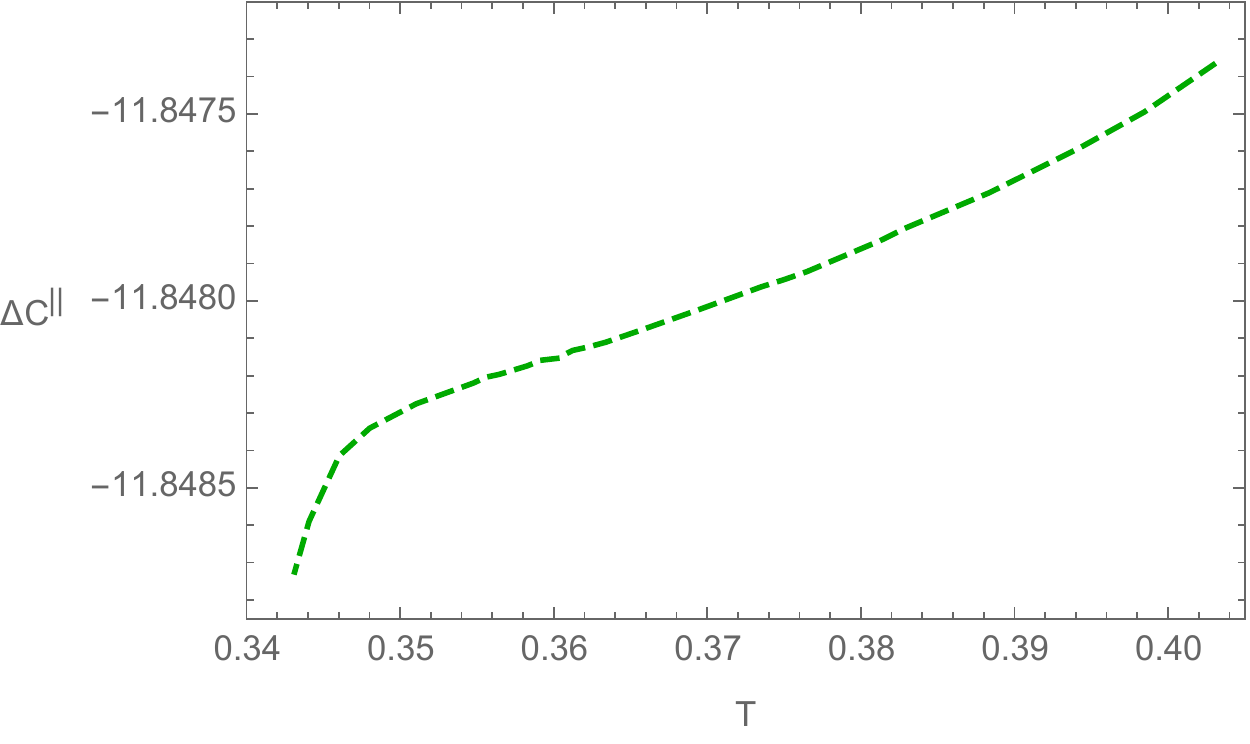}
\includegraphics[width=57 mm]{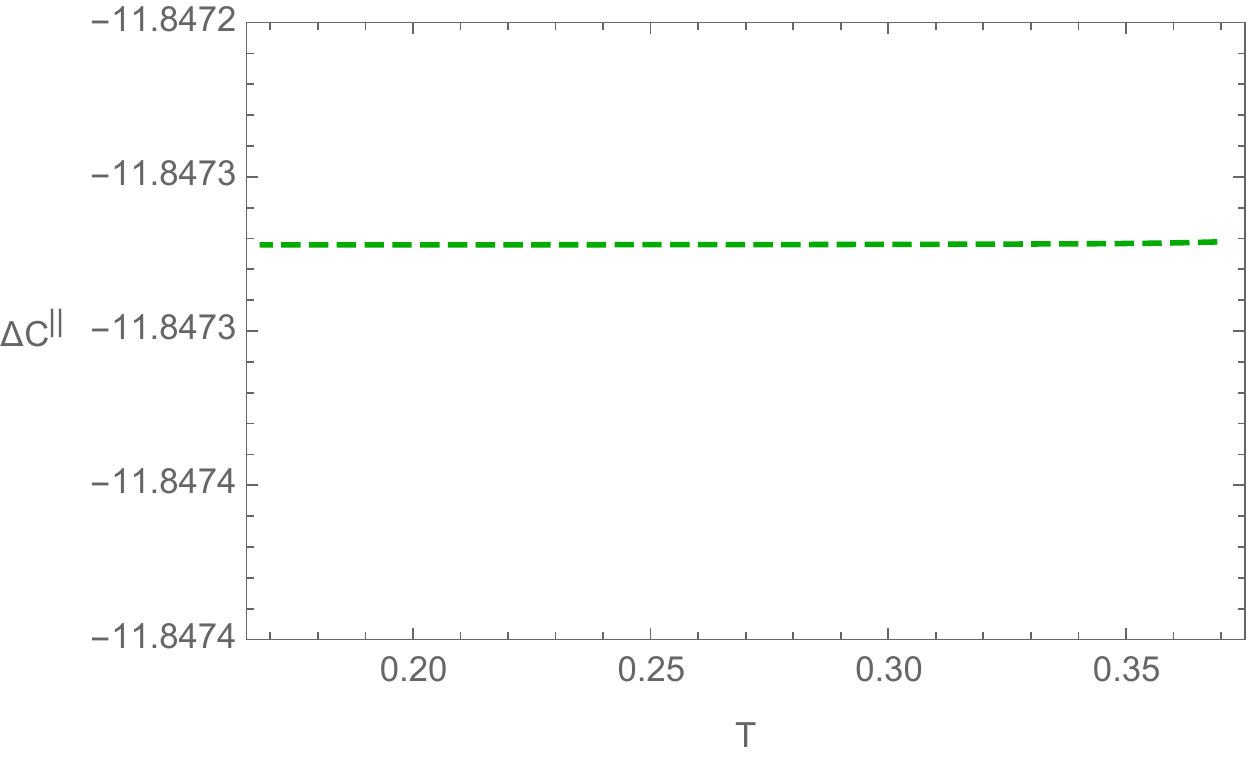}
\caption{In all three panels $l=0.5$ and in the isotropic case $\nu=1$ and in the anisotropic case $\nu=1.01$. $\Delta C^{||}$ as a function of $T$ at $\mu=0$ for the stable branch, that is large black holes ones (left). $\Delta C^{||}$ as a function of $T$ at $\mu=0.05$ for the stable branches i.e. large black holes (middle) and small black holes ones (right).}
\label{fig44}
\end{figure}

In order to study the effect of anisotropic parameter on HSC in our system, we have plotted $\Delta C^{||}\equiv \frac{8\pi RG_{5}}{L^{2}}(\mathcal{C}^{||}-\mathcal{C}_{iso})$ as a function of temperature for stable solutions in Fig \ref{fig44}. The left panel is $\Delta C^{||}$ for stable branch at $\mu=0$, and the middle and right panel are related to the case of $\mu\neq 0$ for stable branches, large black holes (middle) and small black holes (right). It shows $\Delta C^{||}$ is negative which means anisotropy causes decrease of HSC. We want to justify this decrease by saying that the system we study consists of a state and its environment which forms a closed system. In this case, the external source that injects  anisotropy into the boundary field theory is the environment. In the initial situation the source of anisotropy is zero and therefore the information for specifying it is zero too. Thus, the needed information for specifying the whole closed system is only related to the state. However, during the assumed evolution, in the final situation, the external source value becomes non-zero and we need information to specify it. According to the principle of information conservation which states information is never lost from a closed isolated system \cite{information}, we claim with increase of complexity of the external source in the final situation, the complexity of the state must decrease, as you can see from Fig \ref{fig44}. We have proposed the same justification for an external heat source in \cite{co15}.

\begin{figure}[H]
\centering
\includegraphics[width=70 mm]{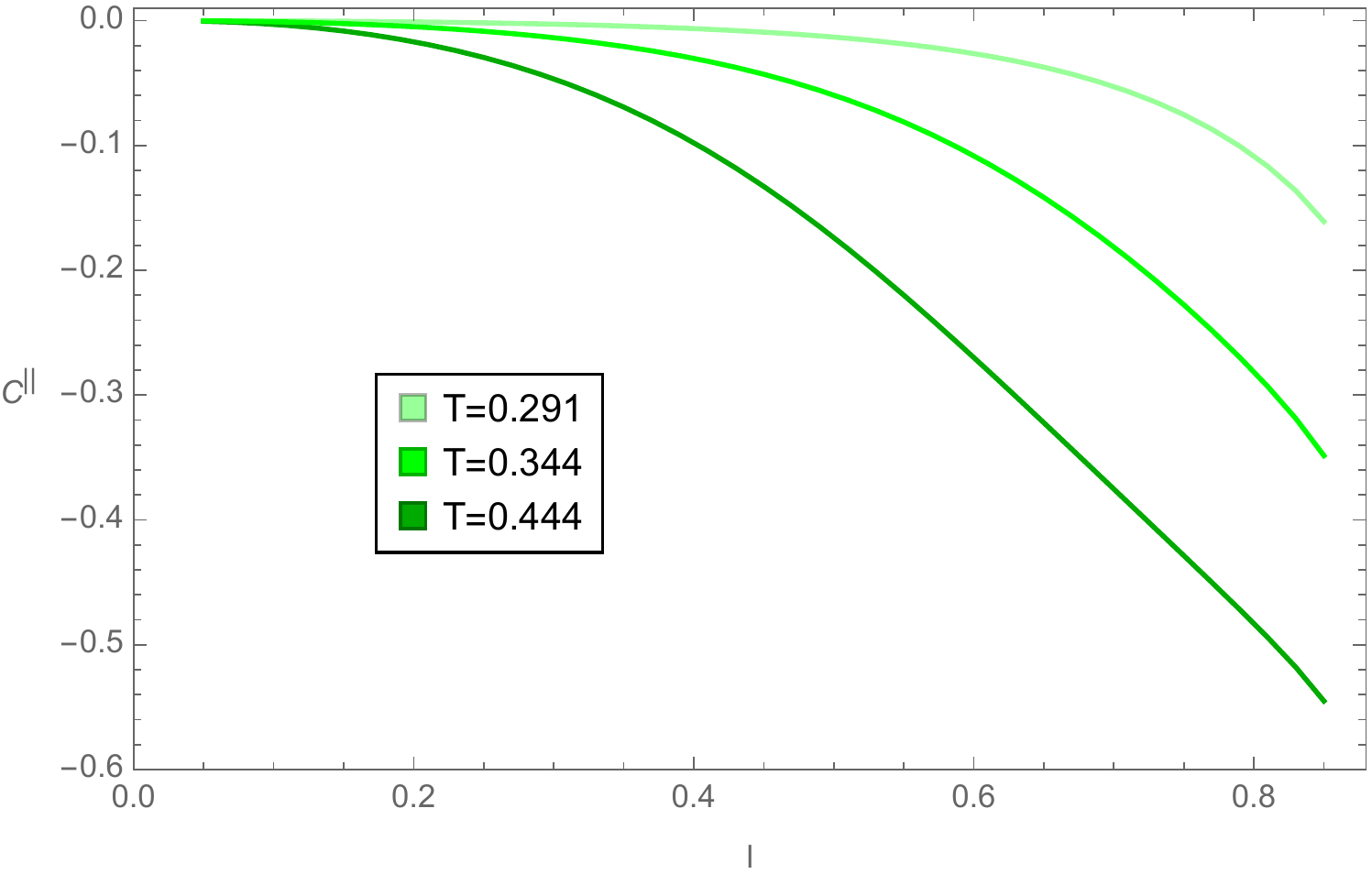}
\includegraphics[width=70 mm]{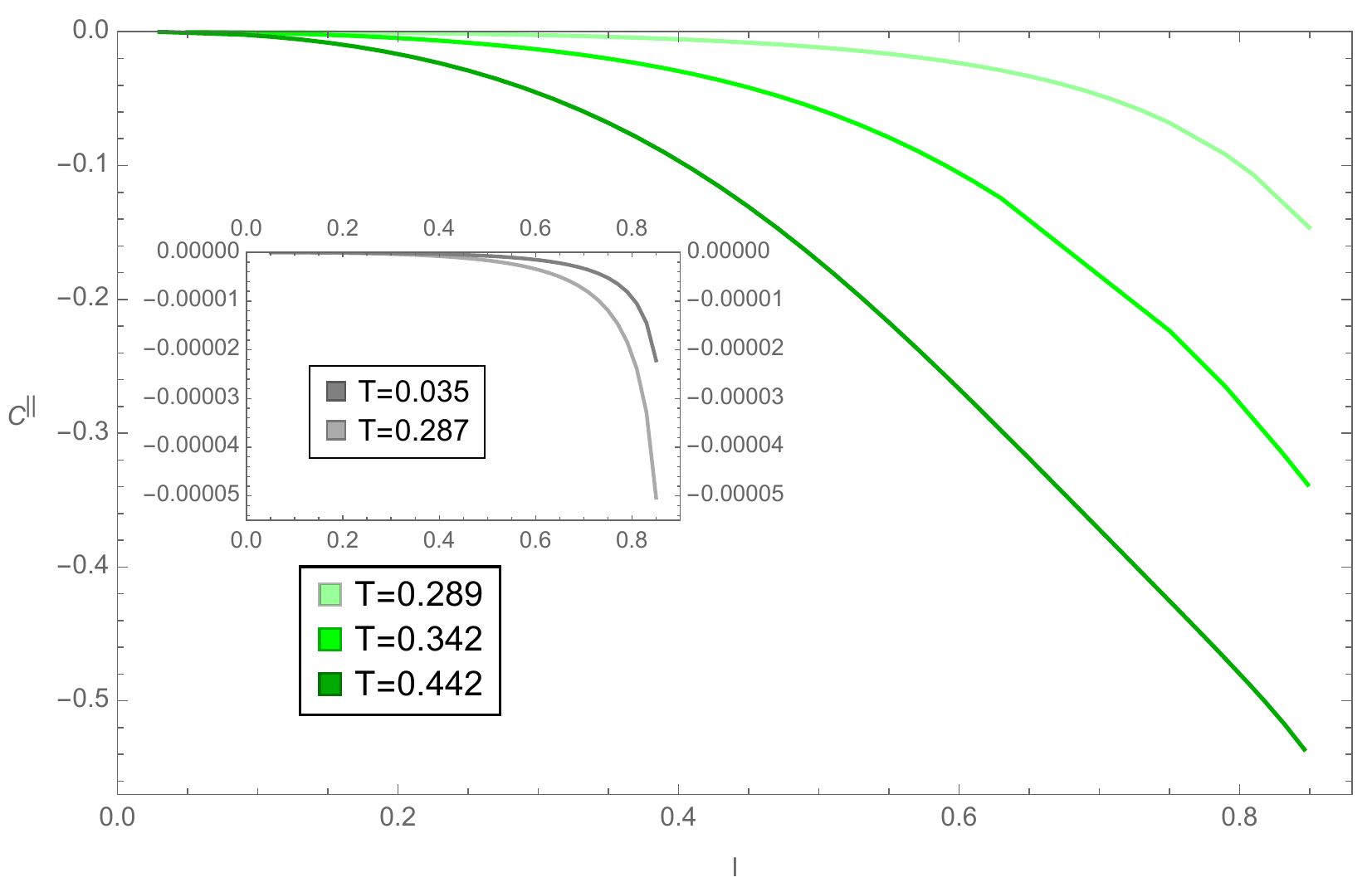}
\caption{Left: $C^{||}$ as a function of $l$ at $\nu=2$ and $\mu=0$ for temperatures above and near $T_{HP}\approx 0.290$. Right: $C^{||}$ as a function of $l$ at $\nu=2$ for $\mu=0.05$ and different values of temperature below and above the phase transition temperature, $T_{BB}\approx 0.288$.}
\label{fig5}
\end{figure}

Fig \ref{fig5} has presented the dependence of $C^{||}$ on subregion length at different values of temperature. The left panel shows $C^{||}$ as a function of $l$ at $\mu=0$ and has been plotted for large black holes i.e. the stable branch. $C^{||}$ is almost independent of temperature for small values of $l$, however with the increase of $l$ the difference is revealed, which compared with the isotropic case, the left panel of Fig \ref{fig3}, this takes place at smaller $l$. The right panel of Fig \ref{fig5} shows $C^{||}$ as a function of $l$ at $\mu=0.05$ for temperatures below (the small black hole branch) and above (the large black hole branch) the phase transition temperature, $T_{BB}$. In this case, $C^{||}$ undergoes a jump when the temperature crosses the $T_{BB}$. 

\subsubsection{Perpendicular case}
\begin{figure}[H]
\centering
\includegraphics[width=70 mm]{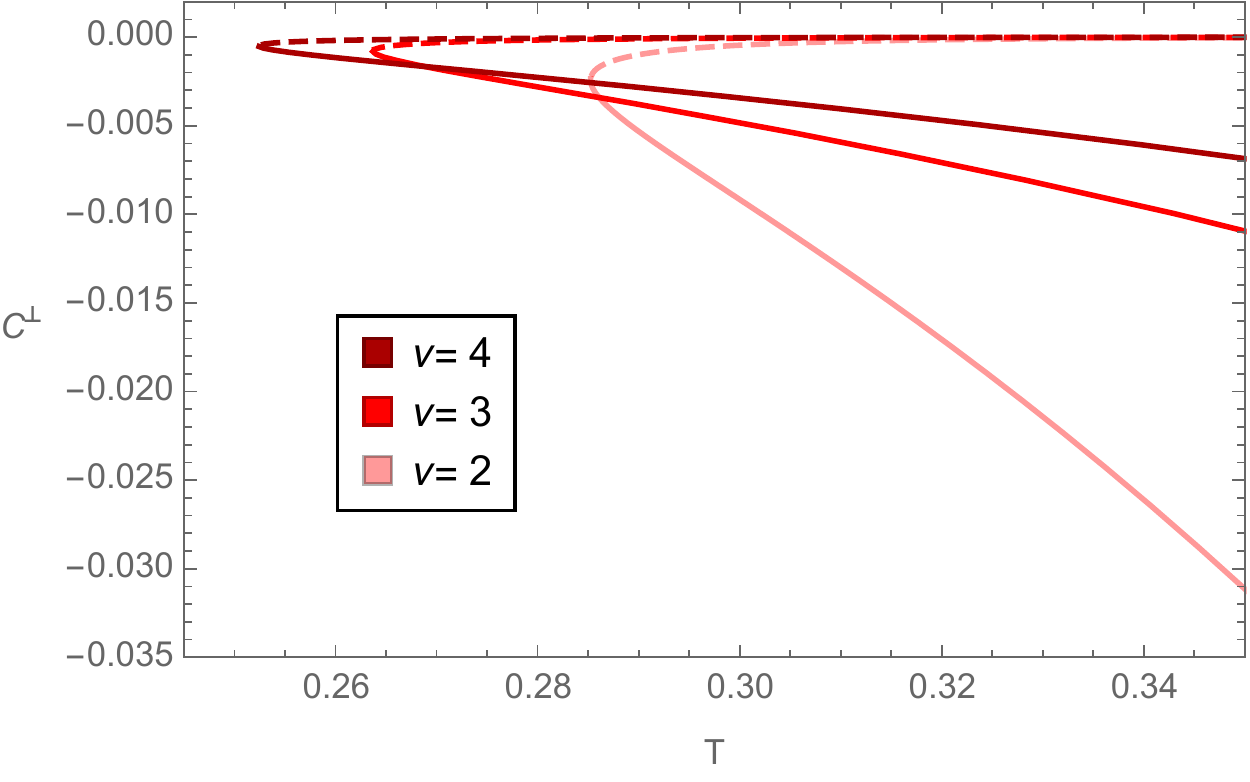}
\includegraphics[width=70 mm]{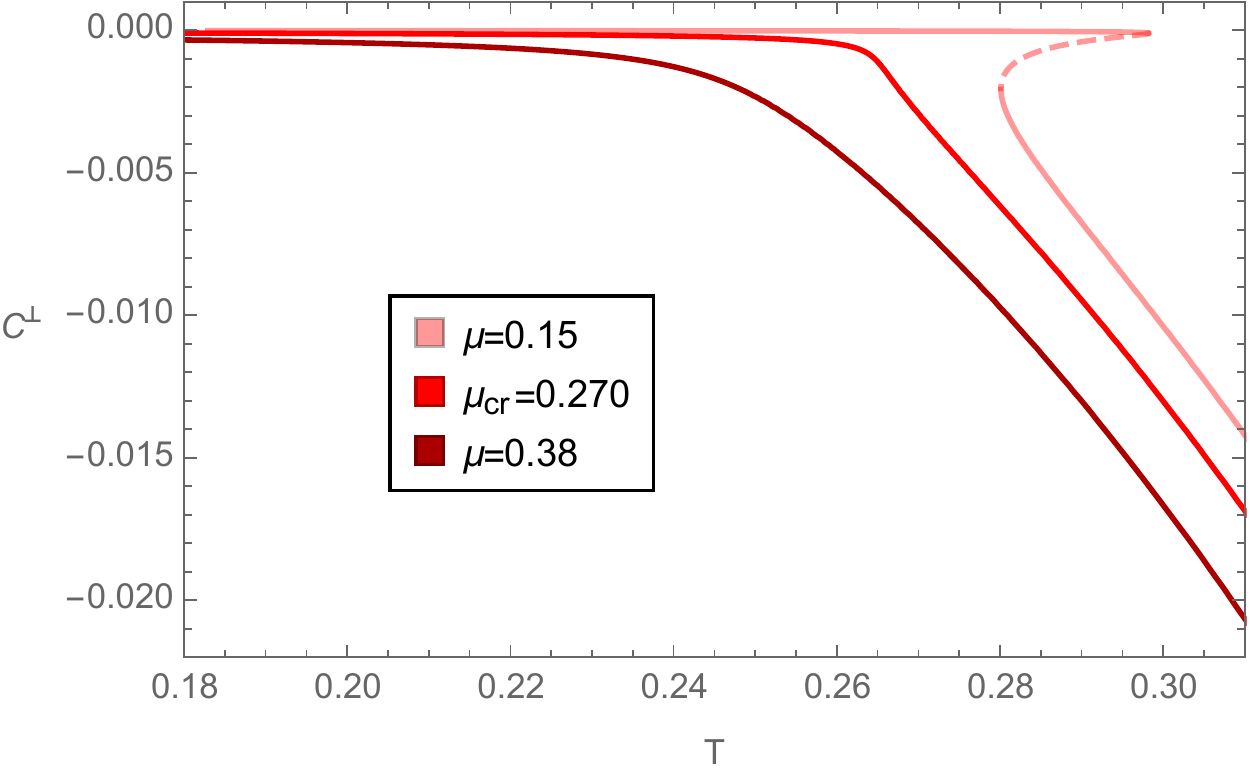}
\caption{Left: $C^{\bot}$ as a function of $T$ for three values of $\nu$ for $l=0.5$ at $\mu=0$. Right: $C^{\bot}$ in terms of $T$ for different values of $\mu$ at $l=0.5$ and $\nu=2$.}
\label{fig7}
\end{figure}

Using \eqref{normal2} for the perpendicular case, we have plotted $C^{\bot}$ in terms of temperature in Fig \ref{fig7}. The left panel shows the case of $\mu=0$ in which $C^{\bot}$ as a function of temperature is a double-valued function for three values of $\nu$ as expected. The same argument of the previous subsections applies here, for specifying the stable and unstable branch. The right panel presents $C^{\bot}$ in terms of temperature as a multivalued function for $\mu<\mu_{cr}$ and a one-value function by passing $\mu_{cr}$. To characterize the stable branches and unstable branches we can use the same arguments as before. 

\begin{figure}[H]
\centering 
\includegraphics[width=57 mm]{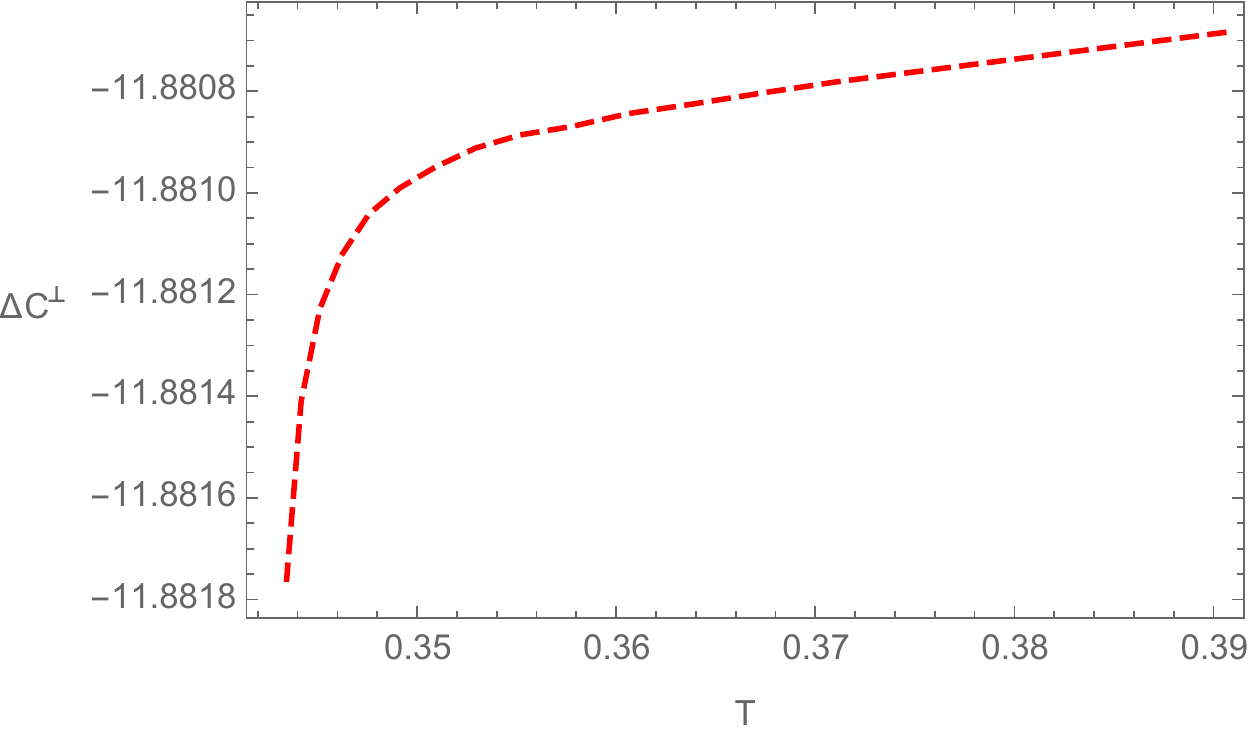}
\includegraphics[width=57 mm]{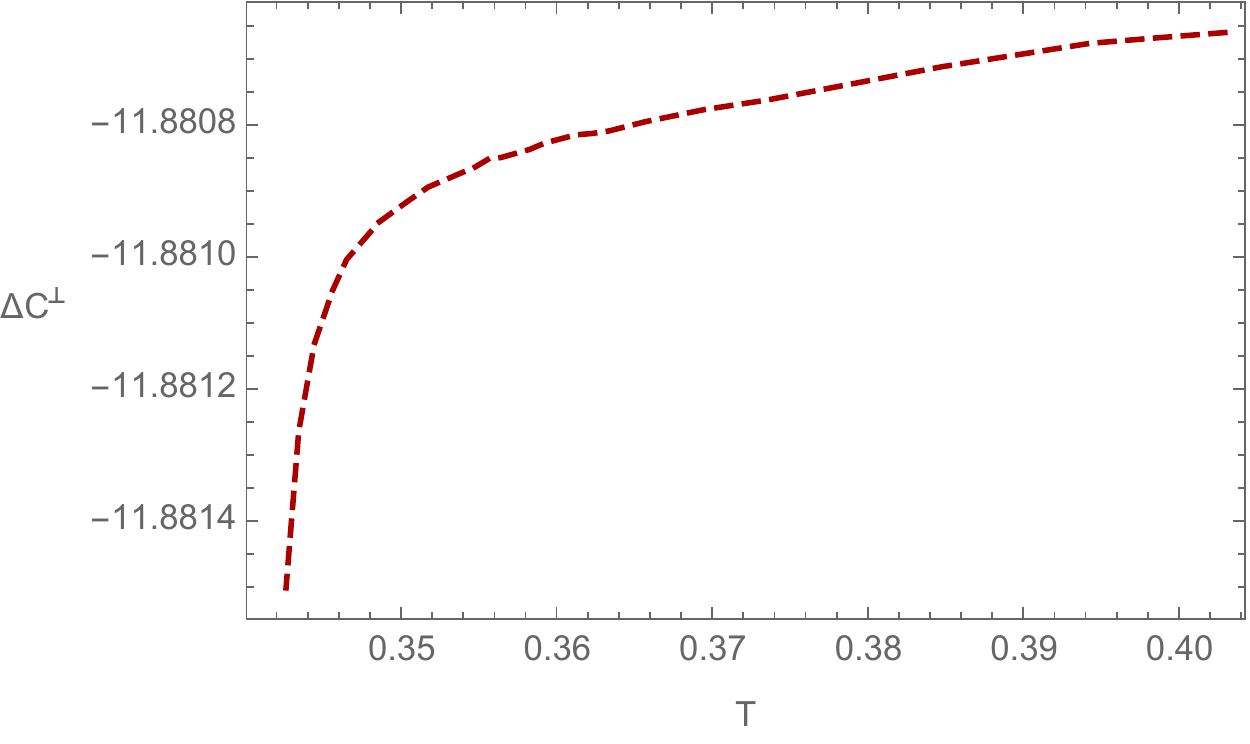}\includegraphics[width=57 mm]{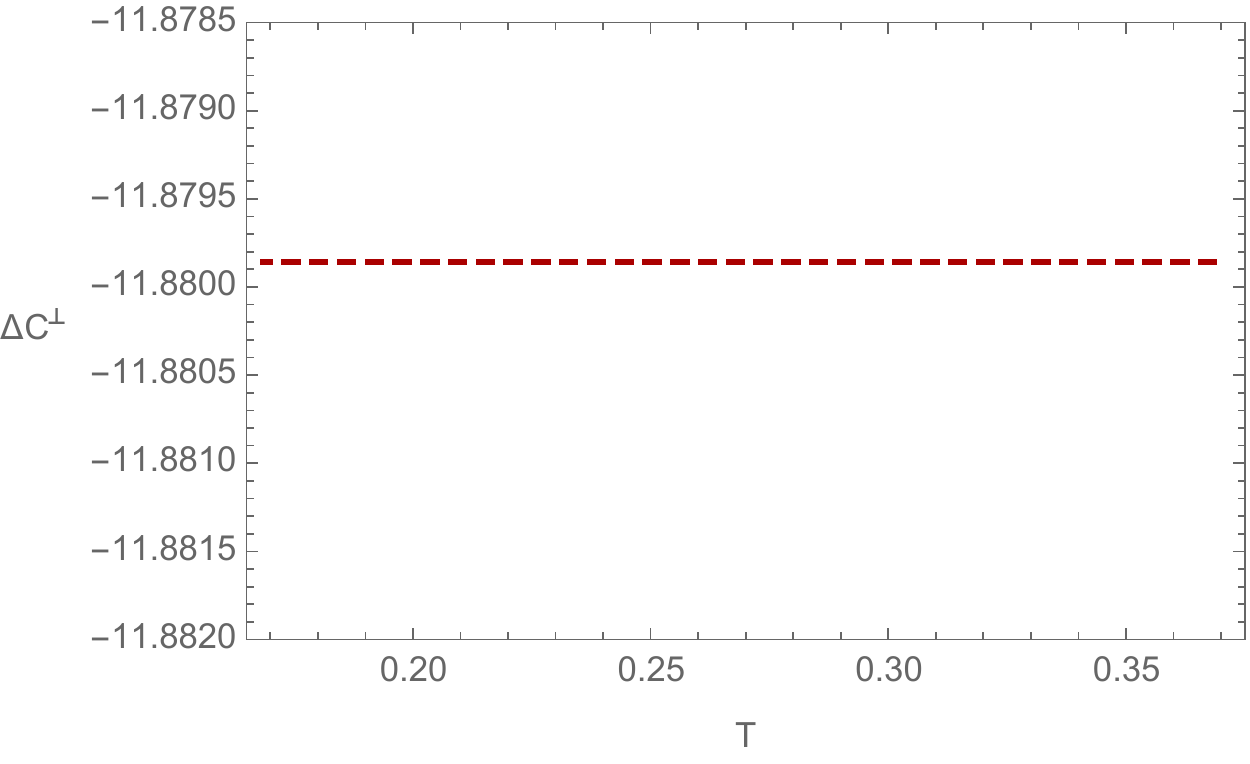}
\caption{In all three panels $l=0.5$ and in the isotropic case $\nu=1$ and in the anisotropic case $\nu=1.01$. $\Delta C^{\bot}$ in terms of $T$ at $\mu=0$ for the stable branch, that is large black holes ones (left). $\Delta C^{\bot}$ as a function of $T$ at $\mu=0.05$ for the stable branches i.e. large black holes (middle) and small black holes ones (right).}
\label{fig77}
\end{figure}

To study the effect of the anisotropic parameter on HSC in this case, we have plotted $\Delta C^{\bot}\equiv \frac{8\pi RG_{5}}{L^{2}}(\mathcal{C}^{\bot}-\mathcal{C}_{iso})$ as a function of temperature for stable solutions in Fig \ref{fig77}. The left panel indicates $\Delta C^{\bot}$ at $\mu=0$ case for stable branch (large black holes), and the middle and right panel shows the case of $\mu\neq 0$ for stable branches, large black holes (middle) and small black holes (right). Like the parallel case, the anisotropy effect on HSC of the state under probe is a decreasing effect. We justify this reduction by considering the whole closed system consists of a state and its environment, just like what was described in detail in the previous section for the parallel case. Another thing that is clear from this Fig compared to Fig \ref{fig44} is that $C^{||}>C^{\bot}$. We have no argument to justify it at this point, but we guess it might be related to the screening effect. In the sense that the effects of anisotropy causes a stronger screening in the perpendicular direction and thus we need less information to specify the state in this case, in agreement with \cite{co14}.

\begin{figure}[H]
\centering
\includegraphics[width=70 mm]{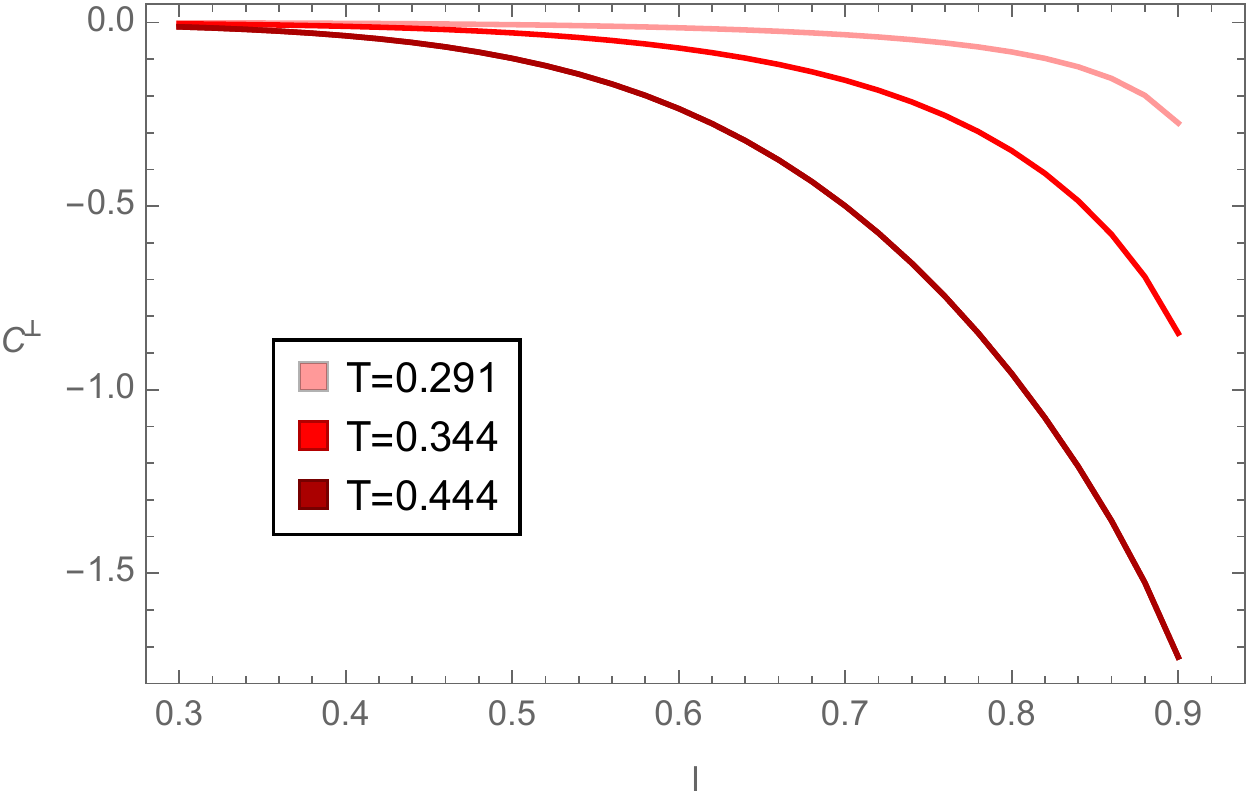}
\includegraphics[width=70 mm]{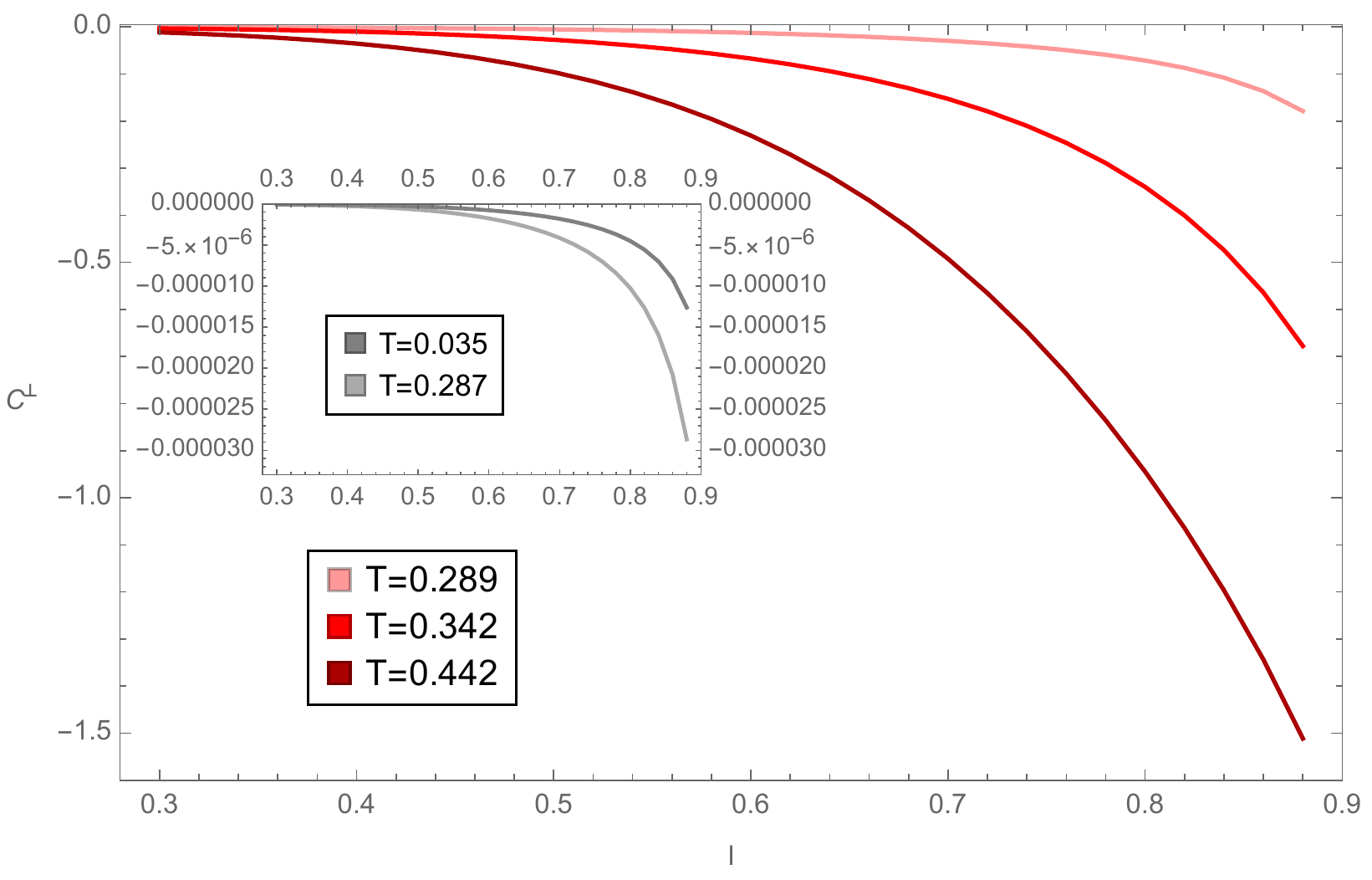}
\caption{Left: $C^{\bot}$ as a function of $l$ at $\nu=2$ and $\mu=0$ for temperatures above and near $T_{HP}\approx 0.290$. Right: $C^{\bot}$ as a function of $l$ at $\nu=2$ for $\mu=0.05$ and different values of temperature below and above the phase transition temperature, $T_{BB}\approx 0.288$.}
\label{fig55}
\end{figure}

Fig \ref{fig55} indicates $C^{\bot}$ as a function of $l$ for different values of temperature. The left panel is the case of $\mu=0$ and has been plotted for stable branch (large black holes) which approach the phase transition temperature, $T_{HP}$ from above. The difference between values of $C^{\bot}$ appears with the increase of $l$, which compared with the isotropic case, this takes place at smaller $l$, like the parallel case. The right panel shows $C^{\bot}$ as a function of $l$ for $\mu=0.05$. This panel has been plotted for temperatures below (the small black holes) and above (the large black holes) the phase transition temperature, $T_{BB}$. In this case, like parallel case, $C^{\bot}$ has a jump when the temperature passes the $T_{BB}$. 

\begin{thebibliography}{99}



\bibitem{CasalderreySolana:2011us}
  J.~Casalderrey-Solana, H.~Liu, D.~Mateos, K.~Rajagopal and U.~A.~Wiedemann,
  ``Gauge/String Duality, Hot QCD and Heavy Ion Collisions,'' Cambridge University Press, 2014.
   \href{http://arxiv.org/abs/1101.0618}{[arXiv:1101.0618 [hep-th]]}.

\bibitem{one}
  R.~Baier, A.~H.~Mueller, D.~Schiff and D.~T. Son, 
  ``Bottom up thermalization in heavy ion collisions,'' Phys.~Lett.~B {\bf 502}, 51 (2001)
   \href{http://arxiv.org/abs/0009237}{[arXiv:0009237 [hep-th]]}.

\bibitem{two}
  M.~Strickland, 
  ``Thermalization and isotropization in heavy-ion collisions,'' Pramana {\bf 84}, 671, (2015)
   \href{http://arxiv.org/abs/1312.2285}{[arXiv:1312.2285 [hep-th]]}.

\bibitem{three}
  D.~Mateos and D.~Trancanelli, 
  ``The anisotropic N=4 super Yang-Mills plasma and its instabilities,'' Phys.Rev.Lett {\bf 107}, 101601, (2011)
   \href{http://arxiv.org/abs/1105.3472}{[arXiv:1105.3472 [hep-th]]}.

\bibitem{four}
  D.~Giataganas and U. Gursoy and J. F. Pedraza, 
  ``Strongly-coupled anisotropic gauge theories and holography,'' Phys.Rev.Lett {\bf 121}, 121601, (2018)
   \href{http://arxiv.org/abs/1708.05691}{[arXiv:1708.05691 [hep-th]]}.

\bibitem{five}
   I.Y. Aref’eva, A.A. Golubtsova, 
  ``Shock waves in Lifshitz-like
spacetimes,'' JHEP {\bf 1504}, 011 (2015)
   \href{http://arxiv.org/abs/1410.4595}{[arXiv:1410.4595 [hep-th]]}.

\bibitem{co1}
 Micheal A. Nielsen and Isaac L. Chuang, ``Quantum Computation and Quantum Infromation,'' Cambridge university press, (2010), 702 p.

\bibitem{co2}
R.~Jefferson and R.~C.~Myers ``Circuit complexity in quantum field theory,'' JHEP  \textbf{10} (2017) 107, 
\href{http://arxiv.org/abs/11707.08570}{[arXiv:1707.08570[quant-ph]]}.

\bibitem{co3}
D. Stanford and L. Susskind, ``Complexity and Shock Wave Geometries,'' Phys. Rev. D \textbf{90}, no.12, 126007 (2014) 
 \href{http://arxiv.org/abs/1406.2678}{[arXiv:1406.2678[hep-th]]}.


\bibitem{co4}
A. R. Brown, D. A. Roberts, L. Susskind, B. Swingle and Y. Zhao, ``Complexity, action, and
black holes,'' Phys. Rev. D \textbf{93}, no.8, 086006 (2016) \href{http://arxiv.org/abs/1512.04993}{[arXiv:1512.04993[hep-th]]}. 

 \bibitem{alishahiha}
  Mohsen~Alishahiha,
  ``Holographic complexity,'' Phys.~Rev.~D.~ {\bf 92}, 126009 (2015)
  \href{http://arxiv.org/abs/1509.06614}{[arXiv:1509.06614[hep-th]]}. 

\bibitem{co5} 
    Omer~Ben-Ami, Dean~Carmi,
  ``On Volumes of Subregions in Holography and Complexity,'' JHEP {\bf 1611}, 129 (2016)
  \href{http://arxiv.org/abs/1609.02514}{[arXiv:1609.02514[hep-th]]}.

\bibitem{co6} 
  S.~J.~Zhang,
  ``Complexity and phase transitions in a holographic QCD model,''
  Nucl.\ Phys.\ B {\bf 929}, 243 (2018)
  \href{http://arxiv.org/abs/1712.07583}{[arXiv:1712.07583[hep-th]]}.
 
  \bibitem{co7} 
  S.~J.~Zhang,
  ``Subregion complexity in holographic thermalization with dS boundary,'' Eur.Phys.J.C {\bf 79} (2019) 8,715
  \href{http://arxiv.org/abs/1905.10605}{[arXiv:1905.10605[hep-th]]}.

\bibitem{co8} 
  Pratim~Roy, Tapobrata~Sarkar, 
  ``On subregion holographic complexity and renormalization group flows,'' Phys.Rev. D {\bf 97}, 086018 (2018)
  \href{http://arxiv.org/abs/1708.05313}{[arXiv:1708.05313[hep-th]]}.
 
 \bibitem{co9} 
  R. Fareghbal and P. Karimi, 
  ``Complexity growth in flat spacetimes,''
  Phys. Rev. D {\bf 98}, no. 4, 046003 (2018)
  \href{http://arxiv.org/abs/1806.07273}{[arXiv:1806.07273[hep-th]]}.

\bibitem{co10} 
 M.Alishahiha, A.Faraji Astaneh, M.R.Mohammadi Mozaffar and A.Mollabashi, 
   ``Complexity Growth with Lifshitz Scaling and Hyperscaling Violation,''
  JHEP {\bf 1807}, 042 (2018)
  \href{http://arxiv.org/abs/1802.06740 }{[arXiv:1802.06740 [hep-th]]}.
 
\bibitem{co11} 
M.~Alishahiha, K.~Babaei Velni and M.~R.~Mohammadi Mozaffar,
  ``Subregion Action and Complexity,'' Phys.Rev.D {\bf 99} (2019) 12, 126016,
\href{http://arxiv.org/abs/1809.06031 }{[arXiv:1809.06031 [hep-th]]}.
 
 \bibitem{co12} 
 M.~asadi,
 ``On volume subregion complexity in non-conformal theories,"
 Eur.Phys.J.C 80 (2020) 7, 681
  \href{http://arxiv.org/abs/2004.11306}{[arXiv:2004.11306[hep-th]]}.

\bibitem{co13} 
 Mahsa Lezgi, Mohammad Ali-Akbari and Mohammad Asadi,
 ``Non-Conformality, Subregion Complexity and Meson Binding," Phys.Rev.D {\bf 104}, 026001 (2021) 
  \href{http://arxiv.org/abs/2011.11625}{[arXiv:2011.11625[hep-th]]}.


\bibitem{co14} 
 Mahsa~Lezgi and Mohammad~Ali-Akbari
 ``Note on holographic subregion complexity and QCD phase transition, "
 Phys. Rev. D {\bf 101}, 026022 (2020)
  \href{http://arxiv.org/abs/1908.01303}{[arXiv:1908.01303[hep-th]]}.

\bibitem{co15}
  Mahsa lezgi and Mohammad Ali-Akbari,
  ``Complexity and uncomplexity during energy injection,'' 
  Phys.~Rev.~D {\bf 103} (2021) 12, 126024
  \href{http://arxiv.org/abs/2103.05023}{[arXiv:2103.05023[hep-th]]}.

 \bibitem{unco}
  Leonard Susskind,
  ``Three Lectures on Complexity and Black Holes,'' 
  \href{http://arxiv.org/abs/1810.11563}{[arXiv:1810.11563[hep-th]]}. 

\bibitem{unco2}
  Adam~R.~Brown and Leonard Susskind,
  ``Second law of quantum complexity,'' 
  Phys.~Rev.~D {\bf 97} (2018) 8, 086015
  \href{http://arxiv.org/abs/1701.01107}{[arXiv:1701.01107[hep-th]]}.  


 \bibitem{ex1}
  Irina Aref'eva, Anastasia A. Golubtsova, Eric Gourgoulhon
  ``Analytic black branes in Lifshitz-like backgrounds and thermalization,'' JHEP~ {\bf 09}, 142 (2016)
  \href{http://arxiv.org/abs/1601.06046}{[arXiv:1601.06046[hep-th]]}. 
  
   \bibitem{ex2}
  Irina Aref'eva,
  ``Holography for Heavy Ions Collisions at LHC and NICA,'' EPJ Web Conf~ {\bf 164}, 01014 (2017)
  \href{http://arxiv.org/abs/1612.08928}{[arXiv:1612.08928[hep-th]]}. 

 \bibitem{metric}
  Irina Aref'eva, Kristina Rannu,
  ``Holographic Anisotropic Background with Confinement-Deconfinement Phase Transition,'' JHEP~ {\bf 05}, 206 (2018)
  \href{http://arxiv.org/abs/1802.05652}{[arXiv:1802.05652[hep-th]]}. 
 
  \bibitem{metric2}
Irina Aref'eva, Kristina Rannu and Pavel Slepov,
  ``Orientation Dependence of Confinement-Deconfinement Phase Transition in Anisotropic Media,'' Phys.Lett.B~ {\bf 792}, 470-475 (2019)
  \href{http://arxiv.org/abs/1808.05596}{[arXiv:1808.05596[hep-th]]}. 
  
  \bibitem{metric3}
  Irina Aref'eva, Alexander Patrushev and Pavel Slepov,
  ``Holographic entanglement entropy in anisotropic background with confinement-deconfinement phase transition,'' JHEP~ {\bf 07}, 043 (2020)
  \href{http://arxiv.org/abs/2003.05847}{[arXiv:2003.05847[hep-th]]}. 
 
  
 

  
    \bibitem{information}
 L. Susskind and J. Lindesay,
  ``An introduction to black holes, information and the string theory revolution: The holographic universe,'' 
  Hackensack, USA: World Scientific (2005) 183 p.
  
  
  
  
  
  
  
  
  
 \end{thebibliography}
\end{document}